\documentstyle[12pt,aaspp4]{article}
\def\etal{{et al.}\ }
\def\eq{eq.$~$}
\def\Sec{\S $~$}

\def\hi {{\rm {HI}}\ }
\def\mgii {{\rm {MgII}}\ }
\def\mgiii {{\rm {MgIII}}\ }
\def\mgiv {{\rm {MgIV}}\ }
\def\heii {{\rm {HeII}}\ }
\def\mpc{\, {\rm{ Mpc}}}
\def\mpch{\, h^{-1}{\rm{ Mpc}}}
\def\kpch{\,h^{-1}{\rm {Kpc}}}
\def\kms{\,{\rm{ km\ s^{-1}}}}
\def\kpc{\,{\rm{ Kpc}}}
\def\lya{{\rm{ Ly}}\alpha}
\def\cm{\,{\rm{ cm}}}

\def\nhi{\N_{\rm{ HI}}}

\def\J{J_{-21}}

\def\vcir{V_{\rm{ cir}}}
\def\rvir{r_{{ v}}}
\def\rmin{r_{\rm{min}}}

\def\sr{{\rm\,sr}}

\def\erg{{\rm\,erg}}
\def\nmgii{N_{\rm{ MgII}}}
\def\nciv{N_{\rm{ CIV}}}
\def\novi{N_{\rm{ OVI}}}
\def\nhi{N_{\rm{ HI}}}
\def\msun{M_\odot}
\def\cm{{\rm {cm}}}

\def\zsun{Z_{\odot}}
\def\etal{et al.$~$}
\def\kelvin{{\rm\,K}}
\def\erg{{\rm\,erg}}

\def\hz{{\rm\,Hz}}

\def\ref{\parskip=0pt\par\noindent\hangindent\parindent
    \parskip =2ex plus .5ex minus .1ex}
\def\gs{\mathrel{\raise1.16pt\hbox{$>$}\kern-7.0pt
\lower3.06pt\hbox{{$\scriptstyle \sim$}}}}
\def\ls{\mathrel{\raise1.16pt\hbox{$<$}\kern-7.0pt
\lower3.06pt\hbox{{$\scriptstyle \sim$}}}}
\def\gtsima{$\; \buildrel > \over \sim \;$}
\def\ltsima{$\; \buildrel < \over \sim \;$}

\def\gsim{\lower.5ex\hbox{\gtsima}}
\def\lsim{\lower.5ex\hbox{\ltsima}}
\newcount\eqnumber
\def\new{{\the\eqnumber}\global\advance\eqnumber by 1}
\def\ref#1{\advance\eqnumber by -#1 \the\eqnumber
     \advance\eqnumber by #1 }
\def\last{\advance\eqnumber by -1 {\the\eqnumber}\advance 
     \eqnumber by 1}
\def\eqnam#1{\xdef#1{\the\eqnumber}}

\begin{document}
\title{GASEOUS GALACTIC HALOS AND QSO ABSORPTION LINE SYSTEMS}
\author {H. J. Mo$^{1,2}$ and J. Miralda-Escud\'e$^{2}$}
\affil {$^1$Max-Planck-Institut f\"ur Astrophysik, 85748 Garching, Germany}
\and
\affil{$^2$Institute for Advanced Study, Princeton, NJ 08540}
\slugcomment{submitted to the Astrophysical Journal}
\received{        }
\accepted{         }

\begin{abstract}
 A model of Lyman limit QSO absorption systems is investigated where 
they are produced in gaseous galactic halos with a two-phase structure:
a hot phase at the halo virial temperature in approximate pressure 
equilibrium, and a cold, photoionized phase in the form of clouds
confined by the pressure of the hot medium, and falling through it to 
the halo center, probably to accrete on a galactic disk. We show that 
the masses of these clouds must be over a relatively narrow range so
that they are stable against gravitational collapse and evaporation. 
These masses, near $10^6 \msun$, also lead to a covering factor near 
unity. The hot phase is required to have a core radius such that its
cooling time in the core is equal to the age of the halo, and the mass
in the cold phase is determined by the rate at which the hot gas cools.
We calculate the number of Lyman limit systems arising in halos of 
different masses in a CDM model and their impact parameters implied by
these assumptions, and find them to be in reasonable agreement with 
observations. The evolution with redshift is correctly reproduced 
at $z\ls 2$, while at higher redshift we predict fewer absorbers than
observed. We argue that this implies that high redshift Lyman
limit systems arise in more extended infall regions. The observed 
low ionization systems such as MgII are well reproduced as arising 
from the photoionized phase, and are mostly in halos around massive 
galaxies, while CIV selected systems are predicted to be more commonly
associated with lower mass galaxies and larger impact parameters. 
Some CIV can also arise from regions at intermediate temperature at 
the boundaries of the clouds. The hot phase may give rise to detectable
absorption lines in OVI, while the column densities predicted 
for other highly ionized species are low and difficult to observe.
\end{abstract}

\keywords {galaxies: formation - quasars: absorption lines} 

\section {Introduction}

  There is now much evidence that metal line absorption systems in QSO
spectra arise from gas in galactic halos, as first proposed by Bahcall
\& Spitzer (1969). Recent imaging and spectroscopic observations can
identify directly, at low redshift, the galaxies associated with the
MgII absorption systems at typical impact parameters of $30 \kpch$
\footnote {here $h=H_0/[100 \kms\mpc ^{-1}]$ and we use
$H_0=50\kms\mpc ^{-1}$ throughout the paper}
(Yanny, York, \& Williams 1990; Bergeron \& Boiss\'e 1991; Bergeron,
Cristiani, \& Shaver 1992; Bechtold \& Ellingson 1992; Steidel 1993,
1995). However, the origin of the ``halo gas'' and its dynamical
properties are not well understood. 

  During the process of galaxy formation, intergalactic gas must
collapse and move inward through the extended dark matter halos that
are observed around present galaxies. A halo of hot gas at the virial
temperature will form as the kinetic energy of the infalling material
is thermalized in shocks. In low mass halos, the cooling time of this
hot halo gas is short compared to the dynamical time if it contains
all the accreted baryons; therefore, the gas must cool until the density
of hot gas
decreases to a level such that the cooling time is similar to the age
of the system. Due to the increased cooling rate as temperatures drop,
a two-phase medium will naturally form in these conditions of rapid
cooling. Clouds photoionized by a radiation background will
be maintained at an equilibrium
temperature $\sim 10^4 \kelvin$, in pressure equilibrium with the hotter
halo gas. These clouds could form from initial inhomogeneities in the
accreting gas (which should be enhanced in repeated cycles of cooling
and shock-heating of the gas moving through the halo),
or from the ram-pressure stripped interstellar medium
of satellite galaxies. Due to their higher density, clouds will fall to
the center of the halo where they can form stars or globular clusters
in spheroids, or accrete in gaseous disks giving rise to spiral
galaxies (Fall \& Rees 1985; Gunn 1982; Kang \etal 1990). 

  Even though our picture of how galaxies form is still
highly incomplete, what is certain is that at some point in the past
gas had to dissipate and move from the outer parts to the inner parts
of galactic halos, and in this process it must have produced absorption
line systems in any quasar line-of-sight intercepting galactic halos.
It is clear that in order to understand galaxy formation we must first
understand the physical state of the gas out of which galaxies form,
and that absorption line systems are an excellent observational probe
to any gas dissipating in halos. Thus, the
recent observational evidence that the high column density absorption
systems at $z\ls 1$ arise in gaseous halos around galaxies
should be a landmark for our understanding of galaxy formation.
A successful theory of galaxy formation must not
only explain the characteristics of luminous parts of galaxies, but it
is equally important that it can account for the absorption line
systems.

  Based on these considerations, we consider in this paper a simple
two-phase model for the gaseous structure of galactic halos.  
Any gas at intermediate temperatures between the halo virial
temperature and the photoionization equilibrium temperature will have
a very short cooling time, and should not be present in large quantities. 
We show that photoionized clouds confined by the pressure of the
hot halo and falling towards the central galaxy can be the origin
of most absorption line systems with column densities
$\nhi \gs 10^{15} - 10^{16} \cm^{-2}$.
Systems of lower column density can arise in
the infall regions of halos around galaxies and smaller collapsing
systems (Cen \etal 1994), although at low redshift some of them 
may also be produced by pressure confined clouds 
at large radius in galactic halos (Mo 1994).
The model we are investigating here is related to the one 
proposed in Mo (1994). That paper has considered
the properties of clouds, and the pressure of a hot phase, 
that are needed to reproduce the observed
Ly$\alpha$ forest lines at low redshifts.
Here we extend this model to high column-density
systems. We also propose a model for the gaseous structure of 
galactic halos so that the density profile of cold
clouds within a halo, and the number of absorption systems
that are produced, can be predicted.

  In \Sec 2 we describe models for the density profiles of the hot and
photoionized components of the gaseous halos. In \Sec 3 we examine
the physical characteristics of the photoionized clouds, and the ranges
of sizes and masses that are allowed by the physical processes that can
form or destroy them. In \Sec 4 we compare the predictions of our
model with the observations of the quasar absorption line systems.
Our results are discussed in \Sec 5.

\section{The Two-Phase Model for Gaseous Galactic Halos}

  In theories of hierarchical gravitational collapse for galaxy
formation, it is usually assumed that luminous galaxies form at the
center of dark matter halos as the gas collapses, is shocked, and cools
(see White 1995 for a review). At any given epoch, halos can be
parameterized by their circular velocity $\vcir$ at the virial radius
$\rvir$. Here, we shall use as the virial radius that which contains an
average density equal to $18\pi^2 \rho_{\rm {crit}} = 178
\rho_{\rm {crit}}$, where $\rho_{\rm {crit}}$ is the
critical density, and we assume $\Omega=1$. This is the density obtained
for a spherical perturbation that virializes at half the radius of
maximum expansion (e.g., Peebles 1980). The mass of the virialized halo
is then $M_{\rm halo} = 24\pi^3\, 
\rho_{\rm {crit}}\, \rvir ^3$; using also $\vcir =
(GM_{\rm halo}/\rvir)^{1/2}$, we have 
$$\rvir = {\vcir\, t\over 2\pi} ,
\eqno(\new)
$$ 
and 
$$M_{\rm halo}={\vcir ^3 t\over 2\pi G} ,
\eqno(\new)
$$
where
$t$ is the age of the universe.

  According to these theories, halos should continuously be merging and
gravitationally accreting gas from the intergalactic medium. 
The accreted new gas should be shock-heated
to the virial temperature.
In halos of low $\vcir$, the cooling time is shorter than the dynamical
time, and much of the gas can cool before it has time
to move to the center of the halo (e.g., White \& Rees 1978). 
However, there must always be some hot gas left in the halo after
rapid cooling, because once a certain fraction of the gas has cooled
the density of the remaining gas will be sufficiently low for the
cooling time to be longer than the dynamical time, and pressure
equilibrium should be established in the hot halo (Fall \& Rees 1985).
The cooled gas should form clouds in
pressure equilibrium with the hot gas and should then fall through the
galactic halo. In the presence of an external ionizing
background (as well as the X-ray emission from the halo and other
ionizing radiation from local stars in the central galaxy),
the clouds will remain photoionized at an equilibrium temperature near
$10^4 \kelvin$.

  In a massive system where the cooling time of the hot gas 
is longer than the dynamical time of the halo, most of
the gas does not cool rapidly. Only in the inner region,
where the density is higher, the gas will radiate fast enough to cool.
The subsequent evolution of the system is then related to the problem
of cooling flows. The standard cooling flow model used to explain the
observed structure of X-ray halos in elliptical galaxies and clusters
of galaxies (see Fabian 1994 for a review) involves the formation of a
multiphase medium, where the hot gas is separated in phases at different
temperatures cooling independently of each other and forming clouds as
they move to the center. The main difficulty of these models is that
they need to assume that all the phases are comoving, when in fact the
densest phases should rapidly fall to the center (e.g., Balbus \& Soker
1989; Loewenstein 1989; Malagoli \etal 1990). Although a thermal
instability may exist in the presence of magnetic fields (Loewenstein
1990; Balbus 1991), it is not clear if relative motions can be prevented
when the overdensities of different phases become large, and if clouds
can ever form in cooling flows. Once formed, clouds should either turn
into stars as they move through the halo, or otherwise coalesce in a
gaseous disk where they are supported by angular momentum. The latter
process should lead to a spiral galaxy, but it clearly does not take
place in the X-ray halos where cooling is observed, which are known
observationally to contain very gas-poor, giant ellipticals.
The only possibility is then that clouds turn into stars in the halo,
but the stellar mass function must be very different from that in normal
galaxies: massive stars must be suppressed in order to avoid a very
luminous young population of stars, which is not observed in the giant
ellipticals in the center of X-ray halos with large cooling rates.

  An alternative possibility is that there is a central adiabatic core
in X-ray halos where energy is injected by a central source.
Tabor \& Binney (1993) have proposed this model for elliptical galaxies,
where the energy may arise from an active galactic nucleus.
In disk galaxies, the energy may arise from supernova explosions.
Observations of ``superbubbles'' above the disk of our galaxy indicate
that some of the energy from supernovae is indeed released into the
halo (Tenorio-Tagle et al. 1990). Simple calculations show that the
total energy released in supernova explosions in normal spiral galaxies
is similar to the thermal energy of all the gas accreted: for a galaxy
like the Milky Way, a total of $10^8$ supernovae can produce $\sim
10^{59}$ ergs of energy, while an accreted baryon mass of $10^{11}\msun$
with a velocity dispersion $\sim 200 \kms$ has a thermal energy of
$10^{59}$ ergs.
When a sufficient amount of hot gas is produced near 
the center, the halo will be globally unstable to convection,
and this will set up an adiabatic core extending to a radius that
should increase with the amount of energy released.
As the accreted gas reaches the adiabatic core, it may cool and form
photoionized clouds that fall through the adiabatic substrate, 
eventually being deposited
in a gaseous galactic disk, or evaporating into the hot gas.
This could be a way to generate a hot component with a large core
radius, while a cold component moving through accretes to the galaxy
disk.

  In the following, we  
consider a simple phenomenological model for these two components of
gaseous galactic halos, which should hopefully
capture the essence of their structure.

\subsection {The hot phase }

  Let us consider the profile of the hot gas in a halo after a time
$t_M$ since the last big merger. In other words, we assume that a
large amount of gas was accreted into the system at the time 
of the merger, and that during the time $t_M$ since then there has 
only been a much lower rate of accretion. We also assume that
the feed-back effect from the central part of the halo does
not alter the properties of gas at large radius.  
We consider first the case where the cooling time at the
virial radius $\rvir$ is longer than $t_M$, and $t_M$ is longer than the
dynamical time. In this case, the gas at large
radius should preserve its initial density after virialization, since
cooling is not very important.
Numerical simulations of the formation of halos show that
the gas profile will approximately follow the mass profile, with a slope
close to isothermal (e.g., Evrard 1990, Navarro, Frenk, \& White
1995); thus, we assume that the density of hot gas is $\rho_h \propto
r^{-2}$ at large radius. 
There will then be a radius $r_c$ ($<\rvir$) where
the cooling time is equal to $t_M$, which we call the cooling radius.

  At $r < r_c$, the density profile must be shallower than isothermal,
because otherwise most of the cooling radiation would be emitted near
the center. 
In cooling flow models, and when the gravitational potential is
important, the core radius of the gas profile
is similar to the cooling radius (Waxman \& Miralda-Escud\'e 1995).
In general, the detailed density profile of the hot gas
depends on the initial conditions of the assumed multiphase gas when it
starts cooling (Nulsen 1986). A narrow initial temperature
distribution will result in a steep gas density profile
($\rho_h \propto r^{-3/2}$ for an isothermal potential well).
This does not result in a good model for galactic halos, since
it produces an X-ray surface brightness much higher than 
observed. A more extended profile for a cooling flow model
can be obtained when the hot
gas forms a multiphase medium with a wide temperature 
distribution. The flattest possible density profile 
occurs when the temperature profile is adiabatic.
An adiabatic temperature profile can also be produced by the presence 
of a central energy source, as described previously.
For simplicity we will adopt a model where the adiabatic halo
extends all the way to the cooling radius $r_c$.     
This gives the flattest core that is 
consistent with the convective stability of
the hot halo, if the profile outside $r_c$ is isothermal.

 In such an adiabatic halo, the pressure of the hot gas $P_h$ is related 
to $\rho_h$ by $P_h\propto \rho_h^{5/3}$. From the
equation of hydrostatic equilibrium in an isothermal sphere
it then follows that the density and temperature profiles are given by
$$
\rho_h(r)=\rho_h(r_c)\left\lbrack 1-{2\over 5}
{\mu \vcir ^2\over kT_h(r_c)} {\rm ln} {r\over r_c}
\right\rbrack ^{3/2} ;
\eqno(\new )
$$
$$
T_h(r)=T_h(r_c)\left\lbrack 1-{2\over 5}
{\mu \vcir ^2\over kT_h(r_c)} {\rm ln} {r\over r_c}
\right\rbrack ,
\eqno(\new )
$$
where $\mu$ is the average mass per particle.
We assume that the temperature
of the hot gas at the cooling radius $r_c$ is equal to the
virial temperature, so that $T_h(r_c)=T_v\equiv \mu \vcir^2/2k$.
The density of hot gas at $r_c$, $\rho_h(r_c)$,
is obtained by requiring the cooling time at $r_c$
be equal to $t_M$: 
\eqnam{\rhoh}
$$
\rho_h (r_c)=
 {5 \mu\, kT_v\over  
2 \Lambda (T_v)t_M } , 
\eqno(\new ) 
$$
where $\Lambda$ is the cooling rate.
To determine $r_c$, we assume 
$$
\rho_h(r_c)={f_g\vcir ^2 \over 4\pi G r_c^2} ,
\eqno(\new )
$$ 
so that the density of hot gas at this radius is a fraction of $f_g$
of the total density of the halo.

   When $r_c>\rvir$, the cooling time at the virial radius is
already short compared to $t_M$, and the fraction of
gas in the hot phase is small. In this case, 
the cooling radius $r_c$ defined by equations (5) and (6) is simply
a parameter rather than a physical cooling radius.
We assume that the residual hot gas at $\rvir$ is still at the 
virial temperature and has a density such that its cooling time 
is equal to $t_M$, so that $T_h(\rvir)=T_v$ and
$\rho_h(\rvir)=(5 \mu\, kT_v)/[2\Lambda (T_v)t_M]$. 
For simplicity we also assume that the hot gas within $\rvir$
has density and temperature profiles similar to those
described by equations (3) and (4), but with $r_c$ replaced by 
$\rvir$. Outside $\rvir$, the gas has typically not been shocked
and not much hot gas should exist there. 

  The time $t_M$ in equation (5) 
corresponds to the interval between major mergers,
since the gas is then heated to a stage from which it starts cooling.
This time depends on the halo mass and the power spectrum; we ignore
this complication here, and assume $t_M = t/(1+ \Omega)$, where $t$ is
the age of the universe. This reproduces an earlier formation of halos
for low $\Omega$ (see also Fig.8 of Lacey \& Cole 1993 for
$\Omega=1$). The result obtained for the cooling radius in terms of the
halo circular velocity is shown in Figure 1 at several redshifts,
assuming $f_g=0.05$ and a metallicity $Z = 0.3 \zsun$ for the cooling
function adopted from Sutherland \& Dopita (1993). 
For typical $L_*$ galaxies ($\vcir \simeq 250 \kms$) at
$z\sim 0$, the cooling radius is near $150 \kpc$, and it
stays almost constant as we increase the halo velocity, up to scales of
rich clusters of galaxies. For smaller galaxies, the cooling radius is
larger and eventually reaches the virial radius; the value of
$\vcir$ at which this occurs is shown by the thick vertical ticks 
in Figure 1. 

  The total surface brightness emitted in soft
X-rays by these hot halos can be computed given
the emissivity per unit volume, $\epsilon_x(r) = \epsilon_0 \rho^2_h(r)
/\mu_e^2$, where $\mu_e$ is the gas mass per electron.
The result is:
\eqnam{\sxprof}
$$
S_x(R) = {2\epsilon_0\, f_g^2\, \vcir ^4\over 64\pi^3 G^2\, 
\mu_e ^2 r_c^3}\,
F(R/r_c)
$$
$$= 1.5\times 10^{-8} \left({f_g\over 0.05} \right)^2
\left( {\vcir\over 200\kms } \right)^4
\left( { 100\kpc \over r_c } \right)^3 
F\left({R\over r_c}\right)
\erg\,\cm^{-2}\sec^{-1}\sr^{-1},
\eqno(\new)
$$
where $R$ is the radius in projection and, for the density profile
given by equation (3), the dimensionless function $F$ is
$ F(x) = \int_0^{1} dz 
\left[1- (2/5) {\rm ln}\, (z^2 + x^2)\right]^{2}$,
where we have neglected the minor contribution from 
the gas outside the adiabatic core.
Taking the characteristic values in equation (\sxprof), and
$\epsilon_0=2\times 10^{-23} \erg\cm^{3}\sec^{-1}$,
(adequate for temperatures $\sim$ few $10^6 \kelvin$ and
$Z \sim 0.1 \zsun$), we find that for halos with $r_c=100\kpc$ the 
surface brightness at
$R= 100 (20) \kpc$ is $S_x = 1.5\times 10^{-8} (4\times 10^{-8})
\erg\cm^{-2}\sec^{-1}\sr^{-1}$, and the total luminosity within $r_c$ is
$L_x \sim 7 \times 10^{40} \erg\sec^{-1}$.

  The value for the surface brightness obtained at radii $R\sim 20 \kpc$
is consistent with the observational upper limits. For example, McCammon
\& Sanders (1984) find that the X-ray surface brightness within $30 \kpc$
around the galaxy M101 (where the disk circular velocity is $\vcir \simeq
200 \kms$ ; Dean \& Davies 1975) is smaller than $\sim 3\times 10^{-8}
\erg\cm^{-2}\sec^{-1}\sr^{-1}$, corresponding to $L_x < 10^{40} \erg
\sec^{-1}$ within the same radius. This upper limit was
determined by subtracting the emission between $30$ and $50 \kpc$ from
the emission within $30 \kpc$, so for a very extended X-ray halo the
central surface brightness could probably be larger. 
For our galaxy, the emission
measure that is derived from the hot gas at radius $\sim 100\kpc$
is $\sim 2\times 10^{-3}\,{\rm pc\, cm}^{-6}$, which is consistent
with the soft X-ray background (e.g. Wang \& McCray 1993).
Despite of this, it is clear that the density profile of the hot gas 
cannot be much steeper than that assumed in our model.
The hot phase in our model (and the large total X-ray luminosity it
produces) is still consistent with the X-ray observations of
normal galaxies only because we have assumed a large core radius for the
hot gas.

  It is possible that $f_g$ is substantially smaller than $0.05$,
and a steeper density profile is then allowed.
However, the value of $f_g$ in our model is also constrained by the
absorption line systems, as will be discussed later
in the paper.

\subsection {The cold phase}

   The accreted gas that does not stay in the hot phase and cools
must fall through the hot halo in the form of photoionized clouds.
We refer to the gas in these clouds as the ``cold phase''.
As discussed before, the rate at which cold gas accumulates 
in a halo is determined by both gas infall and gas cooling. 
The accumulation rate is in 
general not uniform in time: it may be substantially higher
than the average during a big merger when large amount of gas
is accreted and can cool. 
We shall, however, ignore this in our simple model.

In small halos, where 
the cooling time of the hot gas within the virial radius 
is much shorter than the dynamical time, most of the accreted gas
will have cooled.
Since the total gas mass accreted in a halo with circular velocity 
$\vcir$ is $f_g \vcir ^2 \rvir /G$, the total mass of gas that has
been in the cold phase can be written as   
$$M ={f_g\vcir ^2 \rvir \over G}
-\int_0^{\rvir} 4\pi x^2\rho_h(x)\,dx ,
\eqno(\new) $$
where $\rho_h(x)$ is the density of gas in the hot phase,
as discussed in \S2.1. 
For massive halos where the cooling time in their outer region
is longer than the dynamical time, we assume that the infalling
gas is initially shock-heated to the virial temperature, and cold 
gas results from subsequent cooling. 
In this case, the total mass of gas that has been in the cold phase
can approximately be written as
$$M ={f_g\vcir ^2 r_c \over G}
-\int_0^{r_c} 4\pi x^2\rho_h(x)\,dx ,
\eqno(\new) $$
where we have assumed that
the amount of cold gas outside $r_c$ is negligible.

  We now assume that most of the cold gas has been formed near a
radius $r_{min} \equiv {\rm min}(r_v, r_c)$, and that a constant
mass inflow rate of the cold gas is present within $r_{min}$, and
no cold gas is present outside this radius. This approximation is
adequate when there is heated gas in an adiabatic core, since the
accreting gas would then start forming clouds at $r_c$ when it mixes
with the heated gas, and in a cooling flow model where, for a wide
initial temperature distribution, 
most of the gas is deposited near $r_c$. 
The cold gas should form clouds that will fall through the halo,
and it will therefore stay in the cold phase only for a short time,
before forming stars or merging with the disk being formed in the halo
center.
We assume the mass flow rate to be ${\dot {M}}
=M/t_M$, and that the clouds move to the halo center with a constant
velocity $v_c$.
Assuming also spherical symmetry for the gas distribution,
we can write the density of the cold gas as a function 
of the distance $r$ to the halo center as
$$\rho_c (r) = {\dot M \over 4\pi r^2 v_c } =
{ f_g\vcir ^2 \rmin \over 4\pi G\,t_M\, r^2\,
v_c} \left\lbrack 1-{\rmin ^2\over r_c^2}
\int _0^1 x^2 \left(1-{4\over 5}\ln x\right)^{3/2}
\right\rbrack  ~. 
\eqno(\new) $$
In the absence of a hot phase, the cold gas is in free fall and $v_c$
must be of the order of the virial velocity $\vcir$. The friction of hot
gas may cause cold clouds to move at a terminal velocity which is
smaller than $\vcir$. However, since halos on galactic scales have
cooling times that are similar to the dynamical times, the inflow
velocity should not be much smaller than the virial velocity (see \S 3).
The inflow velocity could be much smaller than $\vcir$ in massive
systems, and also if relative motions of different phases could be
prevented by a magnetic field. For simplicity, we will assume $v_c$ to
be a constant throughout a halo and to have a value of the order of
$\vcir$. We will examine the effect of changing the value of $v_c$.

\section{Properties of the Clouds}

  The cold (photoionized) gas component of galactic halos, for which we
have described a model of the average density profile in \S 2.2,
should be in the form of clouds. We consider various physical processes
affecting the formation and destruction of the clouds, and examine the
range of cloud masses that are allowed.

  We model the clouds as spheres of uniform, isothermal photoionized
gas confined by the pressure of the hot medium, with mass
$M_c= M_5\times 10^5 \msun$,
temperature $T_c=T_4\times 10^4\,K$, and pressure $P/k = P_2\times 10^2
\kelvin\,\cm^{-3}$.
We define the quantity:
$$Q\equiv {M_c \over M_{\rm crit}} = M_c\, G^{3/2}\, \left( {P\over 1.4}
\right)^{1/2}\, \left( {kT_c\over \mu} \right)^{-2} =
{M_5\over 5.6\times 10^3}\, {P_2^{1/2}\over T_4^2} ~,
\eqno(\new) $$
where $M_{\rm crit}$ is the critical mass for gravitational collapse of
a cloud that is isothermal and spherical (Spitzer 1978), and $\mu$ is
the average mass per particle. For $Q \ll 1$, the cloud pressure is
uniform, and for $Q>1$ the cloud is unstable to gravitational collapse.
Since the cloud is in pressure equilibrium with a hot medium of density
$\rho_h(r)$ and temperature $T_h$, the cloud radius is
$R_c=(3M_c/4\pi \rho_h)^{1/3}(T_c/T_h)^{1/3}$. It is useful to define
the radius
$$R_h \equiv \left({kT_h\, f_g\over G\rho_h\, \mu }\right)^{1/2} ~.
\eqno(\new) $$
The quantity $(kT_h/\mu)^{1/2}$ is the gas velocity dispersion, while 
$(f_g/G\rho_h)^{1/2}$ should be of the order of the free-fall time if
the hot gas has a fraction $f_g$ of the total mass. Thus, outside the
core of the hot phase, where the gas mass fraction is close to $f_g$,
$R_h$ is similar to the radius $r$ where the cloud is, and as the radius
is decreases below $r_c$, $R_h$ should become larger than $r$ since
it decreases only as $[T_h(r)\rho_h(r)]^{-1/2}$ with radius.
The radius of the cloud can then be written as:
$$R_c = 0.66\, Q^{1/3}\, f_g^{-1/2} (T_c/T_h)\, R_h ~. \eqno(\new) $$
Given the profile of the average density of cold gas in a halo,
$\rho_c(r)$ (see \Sec 2.1), the covering factor $C$ of the clouds at
a given impact parameter $r$ is:
$$
C=\pi R_c^2 l {\rho_c (r)\over M_c} =
1.1\, Q^{-1/3}\, f_g^{1/2}{\rho_c (r)\over \rho_h(r)} {l\over R_h} ~,
\eqno(\new) $$
where $l$ is the typical size of the absorbing region. The covering
factor decreases with $Q$. Since $R_h \gs l$, we see that if the
average density in cold clouds is not much larger than the density of
hot gas, $Q$ needs to be smaller than one for the covering factor to be
unity.

  The solid line in Figure 2a shows the value of $Q$ for which the
covering factor is one, according to equation (\last). We 
take $f_g = 0.05$,
$T_4=1.5$, $v_c=\vcir $ (where $v_c$, defined in \S2.2, is the
velocity at which clouds move towards halo centers)
and radius $r=30\kpch$. The result is
shown for halos at $z=0.5$, with the metallicity of the 
hot gas $Z=0.3\zsun$. 
For halos with small $\vcir$, higher values of $Q$
are sufficient for a covering factor of unity, due to a larger ratio of
$\rho_c / \rho_h$ (see \eq \last). 
The dot-dashed line corresponds to a constant cloud mass of 
$10^5\msun$.
Figure 2b shows the same quantities but at a radius $r=60\kpch$.
At larger radius, $\rho_c/\rho_h$ is smaller and the covering factors
are lower for a given $Q$.

  The observations of metal-line absorption systems indicate that most
lines of sight within $\sim 30 \kpch$ from normal galaxies at $z\simeq
0.5$ show MgII absorption (Steidel 1993), while outside this radius
absorption occurs less often (i.e., the MgII equivalent width is often
below the observational threshold). At the same time, the absorption
line profiles typically show a few subcomponents 
(Petitjean \& Bergeron 1990), so the covering
factor of the clouds within $r\sim 30 \kpch$ must be about a few. This
says that any clouds giving rise to the absorption systems must have
values of $Q$ slightly below the solid line in Fig. 2a.

  Assuming ram pressure to be the main cause of drag force on a cloud,
the terminal velocity $V_t$ of a cloud at $r$ is approximately given by 
$(V_t/\vcir)^2 \sim [\rho_c(r)/\rho_h(r)]/C$. 
Using equations (3) and (10), and assuming $v_c\sim V_t$, we have
$(V_t/\vcir)^3\sim (r_c^2 r_{\rm {min}}/r^2\rvir)/(\pi C)$.
Thus in massive halos where $\rvir\gg r_c$,
$V_t$ can be significantly smaller than $\vcir$ for clouds with $C=1$. 
For example, at $z=0.5$ $V_t$ is smaller than $\vcir$ 
for such clouds at $r=30\kpch$ only in halos with $\vcir\gsim
250\kms$. This means that clouds in galactic-sized halos should
move at a velocity not much smaller than the velocity
dispersions of these halos. Observations show that QSO MgII systems 
usually resolve into
subcomponents with velocity spread typically of $100$-$200\kms$
(see e.g. Petitjean \& Bergeron 1990). Such velocity substructures
are consistent with their being produced by the motions of
clouds in galactic halos.  

  The cloud sizes are limited by various physical processes. Clouds
with $Q>1$ should collapse gravitationally. In addition, sufficiently
massive clouds will be self-shielding, and as the gas becomes neutral
it can cool to the cold phase at $T\simeq 50 \kelvin$ (e.g., Ferrara
\& Field 1994), and can much more easily collapse gravitationally.

  Small values of $Q$ are limited by other processes. First, if the
covering factor is much larger than unity the clouds will collide and
coalesce as they fall through the halo. Small clouds can also be
evaporated by heat conduction. In the absence of magnetic fields, this
is determined by the evaporation parameter (McKee \& Cowie 1977),
which is essentially the ratio of the electron mean-free path to the
cloud radius, and is given by
$$\sigma_{ev}=1.2 \times 10^4 T_h^2\, \mu/(\rho_hR_c) =
2.3\times 10^{-3}\, {V_{200}^4\, R_{h,100}\over Q^{1/3}\, T_4\,
f_g^{1/2}} ~, \eqno(\new) $$
where $V_{200}\equiv \vcir/(200 \kms)$, and
$R_{h,100} \equiv R_h/(100 \kpc)$.
Clouds with $\sigma_{ev}\gs 0.03$ will evaporate. If the evaporation is
not saturated ($\sigma_{ev}\ls 1$), as is true in our model,
the evaporation time is (Cowie \& McKee 1977)
$$
t_{ev} = 2.2\times 10^{10}\, V_{200}^{-5}\, Q^{2/3}\, T_4 ~ {\rm yr} ~.
\eqno(\new) $$
We have plotted as the long-dashed lines in Figure 2 the value of $Q$
where either the cloud is radiatively stabilized ($\sigma_0 < 0.03$), or
the evaporation time is longer than $r/v_c$, so that the cloud will not
evaporate as it moves through the halo over a radius $r$. Figure 2 shows
that clouds above this evaporation line can still be small enough to
have a covering factor larger than unity at an impact parameter of
$30\kpch$, for halos with $\vcir \ls 230\kms$. Such clouds have masses
typically of about $10^5$-$10^6\msun$. However, the maximum covering
factor of clouds surviving evaporation is lower than unity at an impact
parameter twice as large, as seen in Figure 2b. The value of $Q$ for a
constant $C$ decreases with $\vcir$ at $\vcir \gsim 230\kms$, since for
these halos the density of cold gas is determined by the cooling rate of
hot gas (equation 10), which decreases with $\vcir$ (see Fig.1). Of
course, small clouds might be magnetized and thermally insulated from
the hot phase. In this case, clouds of smaller mass and with a larger
covering factor could survive evaporation. This may be needed to prevent 
cold clouds from evaporation in massive halos with $\rvir>r_c$, in order
to have a unity covering factor at an impact parameter of about
$30\kpch$ (Fig.2a).

  Small clouds can also be disrupted by hydrodynamic instability when
they are accelerated to the terminal velocity. Pressure confined clouds
moving at the sound speed of the hot phase
are disrupted on a timescale of the order of the cloud dynamic time,
$R_c\, (\mu/kT_c)^{1/2}$ (e.g., Murray et al. 1993).
The short dashed curves in Fig.2 show the limit of $Q$, above which
clouds will not be disrupted by hydrodynamic instability
as they move through the halo over a radius $r$. It is clear
from the figure that a cloud that is stable against 
heat conduction may still be disrupted by hydrodynamic 
instability.
However, a cloud that is disrupted into smaller lumps should then be
evaporated into the hot gas, which would only increase the cooling rate 
and thereby promote cloud formation in the hot phase.
Thus, if hydrodynamic instabilities disrupt clouds, a steady state
should be reached where, once the covering factor is large enough,
cloud coalescence and growth from the cooling of hot gas can balance
cloud disruption and evaporation. 
This type of equilibrium is found in numerical simulations
of sheet fragmentation (e.g., Anninos \& Norman 1995; see also
Begelmann \& McKee 1990).
As shown in Fig.2a, the covering factor of such clouds
is not very different from unity within a radius
of about $30\kpch$.

\section{Implications for QSO Absorption Line Systems}

\subsection {Relation between HI and MgII systems}

 In this Section we calculate the average column density of HI and other
metal ions at different impact parameters from a halo. To do this we
need, in addition to the parameters that describe the density profiles
of gaseous halos (\S 2) and the properties of clouds (\S 3), the flux of
the UV background $J(\nu)$ and the metallicity of the absorbing gas $Z$.
We write $J(\nu)$ in the form
$$
J(\nu)=\J (z) \times
10^{-21}\left( {\nu \over \nu_{\rm HI} }\right)^{-\alpha} \Theta (\nu) \,
\erg\cm^{-2}\sr^{-1}\hz^{-1}\sec^{-1} ~, \eqno(\new)
$$
where $\nu_{\rm HI}$ is the hydrogen Lyman limit frequency, $\J (z)$ gives
the redshift dependence of $J(\nu)$, and $\Theta(\nu)$ describes
departures of the spectrum from a power-law. The shape of $J(\nu)$
depends on the spectrum of the sources of the ionizing photons, and on
the absorption by intervening material in the absorption systems
themselves. Here, we shall take a representative example of models where
quasars dominate the background (their spectrum is assumed to
approximately follow a power-law joining the observed UV and X-ray
fluxes). A typical resulting spectra (see Miralda-Escud\'e \& Ostriker
1990; Madau 1992) can be represented approximately by taking
$\alpha = 0.5$, $\J (z)=0.5$ for $z>2$ and
$\J (z) = 0.5\times [(1+z)/3]^{2}$ for $z<2$,
and including a break in the spectrum at $\nu _4\equiv 4$ Ryd (due to
continuum absorption by \heii), with $\Theta(\nu<\nu_4)=1$
and $\Theta_4\equiv \Theta(\nu\ge \nu_4)=0.1$.
Notice that the amplitude of
this break is expected to increase with redshift; recent observational
evidence from the strength of the He\thinspace II forest suggests that
the break is much larger at $z=3$ (Jakobsen \etal 1994).
We neglect also
other spectral modifications due to emission from $\lya$ clouds, as
described in Haardt \& Madau (1995).

  Our models can predict the number of absorption systems as a function
of the \hi column density that should be seen in halos of different
$\vcir$, given the density profiles for the cold clouds of \Sec 2, and
given the ionizing radiation background. A comparison with observations
is therefore only possible for systems selected from the
\hi column density. For the range of column densities we are interested
in, these can be detected as Lyman limit systems. Although the number
and evolution of Lyman limit systems has been well studied (Sargent,
Steidel, \& Boksenberg 1989; Storrie-Lombardi \etal 1994;
Stengler-Larrea \etal 1995), it has not yet been possible to study their
relationship to galactic halos owing to the small number of known
systems at low redshift. However, this situation will improve as more
low-redshift quasar spectra are observed with HST (see Bahcall
et al. 1995).

  In the meantime, the relation of absorption systems with galactic
halos has only been done from \mgii selected surveys, where all systems
with a \mgii equivalent width above a certain threshold (usually
$0.3 \AA$) are included. The relation between a \mgii equivalent width
and an \hi column density is complicated and should probably show a
large scatter from system to system. First of all, the column density
of \mgii corresponding to a given equivalent width depends on the
degree of saturation of the lines, and the number of subcomponents. For
an unsaturated \mgii line, we have $W_{\mgii}
\approx 0.3 \AA \, (N_{\mgii}/
10^{13} \cm^{-2})$. The components of the observed \mgii lines are
generally saturated (e.g., Steidel \& Sargent 1992), but for $W < 0.3
\AA$ the lines are more optically thin (see Fig.8 in Petitjean \&
Bergeron 1990). We could then assume a corresponding limit of
$N_{\mgii} > 10^{13}\cm^{-2}$ for the \mgii selected surveys, although
systems with higher column densities but narrow components would not be
included. In addition, the ratio $\nmgii/\nhi$ is proportional to the
metallicity, and depends also on the ionization parameter and
self-shielding in the clouds.

  The observations of the metallicity show a large scatter from system
to system. At high redshift, the values obtained for Lyman limit systems
(LLS) from photoionization models are lower than $\sim 0.1Z_\odot$
(Steidel 1990; Petitjean, Bergeron, \& Puget 1992; Vogel \& Reimers 1994),
and at lower redshift ($z\lsim 1$), the metallicity of MgII systems can
be as high as half the solar value (Bergeron \etal 1995). We will take
$Z=0.3Z_\odot$ and examine the effect of changing $Z$.

  In Figure 3 we show $\nmgii/\nhi$ as a function of the ionization
parameter, $\Gamma$ (defined as the ratio of ionizing photon number
density to the total hydrogen number density), for two values of
the \hi column density through a cloud, and different models of the
photoionizing spectrum, as shown in the figure (see \eq [\last]). We
have used the code CLOUDY (Ferland 1993) to
calculate the relative column densities, which assumes a plane-parallel
geometry. The result is shown for a fixed temperature
$T = 15000 \kelvin$, and for $Z = 0.3 Z_\odot$ (the plotted ratio
scales proportionally to the metallicity). A ratio $\nmgii/\nhi \simeq
10^{-4}$ corresponds to having the same threshold for detecting a
halo in absorption as a Lyman limit system (which requires
$\nhi \gs 10^{17} \cm^{-2}$) and as a \mgii system
with $\nmgii \gsim 10^{13}\cm ^{-2}$. We see that
$\nmgii/\nhi$
depends strongly on the ionization parameter and the spectrum of the
background. For $\Gamma < 10^{-3}$, the ratio increases with ionization
parameter, since MgII can be destroyed by the charge-transfer
reaction with a proton to give MgIII and HI. At
higher ionization parameters, the \mgii column density drops very fast
owing to the ionization of \mgiii to \mgiv. Since the ionization
potential of \mgiii is 80.1eV, this occurs in
the presence of photons above the \heii ionization edge. Thus, the
decline of \mgii with ionization parameter is reduced as the cloud
becomes optically thick, or as the break at the \heii edge is increased
in the spectrum of the background.  Figure 4 shows the ionization parameter ($\Gamma$) as a function of the halo circular velocity, according to our model for the pressure
profile of \Sec 2, for clouds at various radii in halos of different
$\vcir$ at $z=0.5$. The variation of $\Gamma$ with $\vcir$ is very fast
(much larger than the variation with impact parameter), due to the
lower pressure in low $\vcir$ halos.

  We therefore conclude that there is a substantial uncertainty in
transforming the predictions of our model for the Lyman limit systems
into the predictions for the \mgii selected samples of absorption
systems.

\subsection{Lyman limit systems}

  Figure 5 shows the HI column density 
($\nhi$) as a function of the impact parameter ($D$)
for halos of various $\vcir$ at $z=0.5$. The value of $z$
chosen here is similar to the mean redshift of the MgII systems 
for which associated galaxies are identified (Steidel 1995).
We take $f_g=0.05$, $Z=0.3Z_\odot$ and $v_c= \vcir$.
The value of $f_g$ is chosen to match the 
cosmic density of baryons, $\Omega _b=0.0125 h^{-2}$, obtained
from primordial nucleosynthesis (see Walker et al. 1991);
it could be lower if much of the baryonic material accreting in galactic
halos had already formed stars or was in dense gas clumps, instead
of a diffuse halo.
We have used the code CLOUDY to compute the
ionization fraction of hydrogen atoms, 
assuming a plane-parallel slab with $\nhi =10^{17} \cm ^{-2}$.
Because of self shielding, the column density 
may be significantly underestimated for 
$\nhi\gsim 10^{17.5}\cm ^{-2}$; the overestimate
of $\nhi$ at $\nhi< 10^{17}\cm ^{-2}$ is not significant.
The horizontal line shows $\nhi=10^{17} \cm^{-2}$, 
above which a Lyman limit system (hereafter LLS) is produced. 
The HI column density obviously decreases with impact parameter,
because $\nhi$ is proportional to the total column density of cold gas,
which is higher for a smaller
impact parameter, and to the pressure (since neutral fraction 
increases with pressure), which is higher near the center. 
The figure shows that the impact parameter for producing a LLS is 
about $40\kpch$ at $\vcir\sim 200\kms$. 
In halos with $\vcir\gsim 250\kms$, this impact parameter is not much 
larger because the cooling radius is reached, and we assume that no 
clouds are formed outside the cooling radius.

In Figure 6 we show the impact parameter $D$ where
$\nhi =10^{17} \cm ^{-2}$ as a function of 
$\vcir$, for halos at $z=0.5$ 
and various choices of model parameters.
For the same reasons mentioned above, the impact parameter increases
rapidly at low $\vcir$ and then reaches an approximately constant value,
near the cooling radius.
The impact parameter
decreases with the metallicity $Z$ in halos where $\rvir <r_c$ (i.e.,
when the global cooling time is short), 
but increases with $Z$ when $\rvir >r_c$. In the first case,
a lower cooling rate gives a larger pressure in the hot halo
and a larger neutral fraction of hydrogen in the cold clouds.
Since in this case the accretion rate of cold gas is
determined mainly by gas infall rather than gas cooling
(equation 8), the total amount of cold gas that can be accreted
in the halo does not depend on $Z$.
The situation is different when $\rvir >r_c$. In this case,
the accretion rate of cold gas is determined by gas cooling, and
$D$ increases with $Z$, because a larger $Z$
means more gas that can cool (equation 9). 
The dependence of $D$ on $f_g$ is strong, due to the higher pressure when
$f_g$ is larger:
$D$ is reduced by a factor of about 2 when $f_g$ decreases
by the same factor.

  The above result of impact parameter as a function of $\vcir$ is
qualitatively the same as what is observed in MgII systems (Steidel
1993, 1995; Bechtold \& Ellingson 1992). The observation shows that the
maximum impact parameters of their MgII systems to the host galaxies
increase slowly with the K-band luminosities of the host galaxies
($L_K$) at $L_K\gsim 0.1 L_K^*$, and decrease rapidly with $L_K$ at
smaller $L_K$. Unfortunately, similar observations for Lyman limit
systems have not yet been done.

  When the comoving number density of halos 
is known, we can also calculate the 
statistical properties of the absorption systems
(e.g., Mo, Miralda-Escud\'e, \& Rees 1993). 
Here we use the Press-Schechter model (Press \& Schechter
1974) to estimate the number density of halos. As an illustration,
we consider the standard cold dark matter (CDM) 
model with $\Omega _0=1$ and
$h=0.5$. We use the fit in equation (G3) of Bardeen et al. (1986)
for the power spectrum, and normalize it so that
$\sigma_8=0.67$, where $\sigma _8$ is the linear
RMS mass fluctuation in a top-hat window with a radius  
of $8\mpch$.

 Let us first examine the contribution to the total
absorption cross section from halos with different $\vcir$.
To do this we consider the cumulative number of absorption systems
(with $\nhi\ge N_0$) per unit redshift at $z$, produced in halos
with $\vcir \le V_0$:
$$
{{\rm d} N\over {\rm d}z} (z,N_0,V_0)
={c\over H_0}(1+z)^{1/2}\int_0^{V_0}\pi R^2(\vcir,z, N_0)
n(\vcir, z) {\rm d}\vcir,
\eqno(\new)
$$
where $n(\vcir,z){\rm d}\vcir$ is the differential comoving number
density of halos at redshift $z$ and with circular velocity $\vcir$;
$\pi R^2(\vcir, z, N_0)$ is the cross section of a halo
at redshift $z$, and with circular velocity $\vcir$, to produce
an absorption system with $\nhi\ge N_0$.
Figure 7 shows ${\rm d}N/{\rm d}z$ as a function of $V_0$,
for $z=0.5$ and $N_0=10^{17}\cm ^{-2}$. 
In the calculation, 
we have assumed that the covering factor of absorbers
within the maximum impact parameters shown in Fig.6 is
equal to or larger than unity. Thus, the contribution
from massive halos, where the covering factor should be below unity
(see Fig. 2), may be overestimated. From the figure we see that
most of the absorption cross section is produced 
in halos with $\vcir=200$-$300\kms$.
The median value of $\vcir$, below which half of the
systems are produced, is about $250\kms$.
Small halos with $\vcir\lsim 150\kms$ contribute 
only a small fraction of the total cross section.
Thus, our model predicts that most LLSs 
at low redshifts are associated with halos of
large galaxies.

  In fact, the reason why we do not have a much larger contribution
from even larger halos, corresponding to relatively poor clusters of
galaxies, is because we assume that no clouds are produced outside the
cooling radius. We have checked that if a small fraction of the hot
gas (of the order of the ratio of the Hubble time to the cooling time)
is in cold clouds at large radius as well, then more massive halos
(with $\vcir \sim 500 \kms$) dominate the cross section
at low redshifts, and cause the number of LLS to increase very rapidly
with time at present (due to the increasing number of massive halos).
This would be clearly inconsistent with observations.
Therefore, the data on the absorption systems imply that if any
photoionized clouds existed in the outer regions of massive halos,
where the cooling time is long, they should have a small covering
factor.

  The total number of HI absorption systems per unit redshift, 
${\rm d} N/{\rm d}z$, contributed by all halos with any circular
velocity is shown in Figure 8 as a function of
$z$ for systems with $\nhi\ge 10^{17}\cm ^{-2}$.
The data points are observational results
for LLSs, adopted from Stengler-Larrea et al (1995).
Here again we have assumed a unity covering factor
within the maximum impact parameter.
For $z\lsim 2$ our model prediction is consistent with
observation, provided $f_g$ is not much smaller than 0.05.  
This shows that most of the LLSs
at $z\lsim 2$ can be produced by clouds in galactic
halos.

  However, at $z\gsim 3$ our model predicts only about one third
of the LLSs observed.
The reason for this discrepancy is
that the virial radius of a halo large enough to produce Lyman limit
systems at these high redshifts is small [c.f. equation 1;
note also that at $z=3$, all halos with
$\vcir\lsim 300\kms$ have $\rvir <r_c$ (see Fig.1)], and their total
cross section is simply not big enough.
To see this more clearly, we note that at $z=3$ the comoving
number density of halos with $\vcir\ge 150\kms$ is about
$0.13(\mpch)^{-3}$ in the CDM model we are considering. From 
equation (1) we see that the virial radius for 
$\vcir=150\kms$ is about $20\kpch$ at $z=3$. 
Thus, even if each halo with $\vcir\gsim 150\kms$ produces a Lyman
limit system everywhere within its virial radius,
the total number of systems per unit redshift in one line of sight
is only about 1, which is still smaller than the observed
value for the LLSs at $z=3$ by a factor of about 2 to 3
(see Fig.8).
It is also unlikely for
the LLSs at high redshifts to be produced by photoionized
clouds in smaller halos. To show this we recall that
the average gas density within the virial radius
of a halo is ${\overline {\rho_{\rm
{v}}}}=3f_g\vcir^2/(4\pi\, G\rvir^2)$. Assuming ionization
equilibrium we can calculate the average number density
of neutral hydrogen atoms $n_{\rm {HI}}$, and it follows
that the typical HI column density produced
by a halo with circular velosity $\vcir$ is
$$
\nhi\equiv n_{\rm {HI}}\rvir
\sim {1.0}\times 10^{16}\cm^{-2}\,
\left({f_g h^2\over 0.0125}\right)^2
\left({1+z\over 4}\right)^{9/2}
{(\vcir/ 100\kms)\over
hJ_{-21}T_4^{3/4}\phi},
\eqno(\new)
$$
where $\phi$ is the volume filling factor of photoionized
gas. Thus, to produce a LLS with $\nhi\gsim 10^{17}\cm^{-2}$,
the photoionized gas needs to be compressed by a factor
larger than about 10 (so that $\phi\lsim 0.1$).
The compression due to the hot medium, which has 
density given by the cooling condition (equation 5),
is typically $(T_h/T_c)(\rvir/r_c)^2$, where $T_c$
is the temperature of the cold phase. For halos
with $\vcir\lsim 150\kms$ at $z=3$, 
$(\rvir/r_c)\lsim 0.3$ (see Fig. 1)
and $(T_h/T_c)\lsim 50$, giving a compression factor
of about 5 that is still too small to produce a LLS. 
In addition, the observations of metal line profiles show that
the velocity dispersion of the individual components is near
$100 \kms$, indicating also that these absorption systems do not
arise in very small halos.

  We find that if the amplitude $\sigma_8$ of the CDM spectrum
changes from 0.67 (our standard value in the models of Fig. 8) to
1, the number of LLSs at $z=3$ increases by a factor of $1.6$, and
if we then also decrease the intensity of the ionizing background
by a factor of 3, the total change in the number of LLSs is a factor
of $2.5$, bringing them close to what is observed. The predicted
evolution is still in the opposite direction to observed, with the
number of LLSs decreasing at high redshift. However, this  could be
due to a decreasing intensity of the ionizing background at $z > 4$.
The number of observed quasars does not allow $J_{-21}$ to be less than
$0.2$ at $z=3$ (e.g., Haardt \& Madau 1995), but it could be lower at
higher redshift.
 
  Another possibility is that a large fraction of the observed LLSs at
high redshift are produced by photoionized gas outside the virialized
regions. Recent hydrodynamic simulations show that low
column density absorption systems can be produced by gas
in filamentary structures that arise from the gravitational collapse
of initial density fluctuations giving rise to galaxies (e.g., Cen
\etal 1994, Hernquist \etal 1996, Haehnelt, Steinmetz \& Rauch 1996). 
The abundance of LLSs predicted by the
SPH simulations (including an approximate treatment for self-shielding
effects) was found to be substantially lower than observed, by a factor
$\sim 10$, in Katz \etal (1996).
These small number of LLSs in the simulation are produced near very high
density clumps of neutral hydrogen, so few of them arise
outside what we have considered here as the virialized regions in halos.
The number of systems arising in infalling regions would be increased
if small clouds confined by pressure are also present in the
filamentary structures found in the simulations; however, it may
be difficult to obtain high enough column densities from such clouds,
for the same reason why it is difficult to produce them in low-mass
virialized halos: the pressure in the infall regions is low, so the
photoionized gas cannot be confined to a very small volume filling
factor.
   
 This shows that in order to reproduce the observed number of LLSs,
a large fraction of the mass accreting in halos needs to be baryonic,
in the form of small gas clouds having a large covering factor.
The fraction of $5\%$ used in our models is a lower limit,
because the number of predicted LLSs for this fraction is only as large
as observed for models having the largest abundances of massive halos
and the lowest values of the intensity of the ionizing background
allowed by observations of quasars. The conclusion that the minimum
baryon content of the clouds is larger is similar to the result obtained
from the simulations of the lower column density $\lya$ forest
(see Hernquist \etal 1996; Miralda-Escud\'e \etal 1995).

\subsection {MgII systems}

 In the last subsection we have examined the rate of incidence and
impact parameters for LLSs in our model. We now extend the calculation
to MgII systems, and discuss the uncertainties in $J(\nu)$ and $Z$, to
compare with the observations that have been done so far where the
absorption systems are selected from their MgII lines.

 Figure 9 shows $\nmgii$ as a function of impact parameter
for two different shapes of $J(\nu)$.
One has $\alpha=0.5$ without any break;
the other has the same power index but with a break of a factor of 50
at $\nu=4$ Ryd. As we can see from Fig.3, these two cases
represent the two extremes. We assume a metallicity $Z=0.3 Z_\odot$,
and redshift $z=0.5$.
Comparing Fig.9 with Fig.5, we see that the shape
of the MgII column density profile is similar to that
of HI, and the difference can easily be
understood by using Fig.3. The MgII column density is, with respect
to the HI column density, suppressed
in halos with $\vcir\gsim 200\kms$, because $\nmgii/\nhi$ 
increases with $\Gamma$ for $\Gamma\lsim 10^{-3}$.
In halos with $\vcir\lsim 150\kms$, where $\Gamma$ is large,
the suppression of $\nmgii$ is more significant 
for a harder UV flux spectrum (Fig. 9a). 

The observed samples of MgII absorption lines select systems where the
MgII line is above a threshold of equivalent width. Because the lines
are usually saturated, the relation to the column density depends on how
the gas along the line of sight is distributed in clouds at different
velocities: larger equivalent widths will be produced for higher
covering factors of the clouds at a fixed column density.
Here, we shall present results for the number of absorbers
and the impact parameter distribution assuming a fixed threshold
$\nmgii = 2\times 10^{13}\cm ^{-2}$ (reasonable for $W = 0.3 \AA$; 
see e.g., Fig. 8 of Petitjean \& Bergeron 1990), although we should bear
in mind that the average cloud covering factor can also affect the
observed equivalent widths.
Figure 10 shows the results for the impact parameter in halos of
different $\vcir$, at redshift $z=0.5$.
The UV flux spectrum used here has parameters 
$J_{-21}=0.1$, $\alpha=0.5$ and $\Theta (\nu\ge\nu_4) =0.1$ 
in equation (18). 
Comparing Fig.10 with Fig.6, we see that $D$
depends on $\vcir$ in a similar manner for both MgII
and HI systems. At $\vcir\gsim 150 \kms$,
the dependence is weaker for MgII systems, because 
the ratio $\nmgii/\nhi$ increases with $\Gamma$ at 
$\Gamma\lsim 10^{-3}$ (see Fig.3 and \S 4.1). 
At $\vcir\lsim 150\kms$,
$D$ decreases rapidly with $\vcir$, due to the faster cooling rate
and the lower pressures that are implied.
The details of the $D$-$\vcir$ relation depend on the 
shape of the UV flux $J(\nu)$, and on whether or not the
systems are optically thick. The decrease of 
$D$ with $\vcir$ at small $\vcir$ is slower when the UV flux
is softer or as the clouds become more optically thick
(Fig.3). However, the main feature in the dependence of
$D$ on $\vcir$ does not change for all the choices
of $\nmgii/\nhi$ shown in Fig.3.

The above result of impact parameter as a function of $\vcir$
is similar to the observational result of Steidel et al. 
(see Steidel 1995). The observation shows
that the maximum impact parameter $D$ of their MgII systems
to the absorbing galaxies (having typically $L_K\gsim 0.1 L_K^*$)
increases slowly with the K-band luminosity $L_K$ as 
$D\propto L_K^{0.15}$. Assuming Tully-Fisher relation
$L_K\propto \vcir^4$, we have $D\propto \vcir^{0.6}$, which is
shown in Fig.10 by the dotted line. We see  
that such a weak increase of $D$ with $\vcir$ 
can be accommodated in our model, arising from the value of the cooling
radius and the derived gas pressures in different halos.
The very small impact parameters in low $\vcir$ halos agrees with the fact
that MgII absorption systems are not 
commonly found in galaxies fainter than about $0.1L^*$ 
(Steidel 1995).
Our model assumes that the clouds only exist within
the cooling radius, and in practice the number of clouds should
decrease with radius less sharply near the cooling radius.
The increase of $D$ with $\vcir$ for massive halos could,
therefore, be stronger. However, the number of absorption systems
at large impact parameters can also be reduced,
because the covering factor of clouds 
that can survive evaporation in the massive halos giving rise to the
absorption systems at large impact parameter can be lower than
unity (\S 3). 

 To see how different halos contribute to the MgII absorption,
we plot in Figure 11 the cumulative number of MgII systems per unit
redshift at $z=0.5$, produced in halos with $\vcir \le V_0$.
The figure shows that most MgII systems
are produced in halos with $\vcir=150$-$300\kms$.
The median $\vcir$ is about $200\kms$, which is 
slightly smaller than that for the LLSs (\S 4.2), because
the ratio $\nmgii/\nhi$ is reduced in massive halos
due to charge transfer (Fig.3). 
The contribution to the total cross section made by
halos with $\vcir \lsim 150\kms$ is small, which again suggests
that MgII systems should not be commonly found in the halos of
dwarf galaxies.    

  In Figure 12, we show the total number of \mgii systems and the
evolution with redshift, compared to the number of Lyman limit systems,
for two values of the metallicity. The relative number of \mgii systems
to LLSs is correctly predicted when the metallicity is between
$Z= 0.1$-0.3, (compared with the observational results 
in Steidel \& Sargent 1992),
and the evolution is also as observed if the metallicity in these halo
clouds does not decrease very fast with redshift.

\subsection {CIV systems}

  CIV is another commonly found species in QSO absorption line systems
which is a high ionization species, as opposed to MgII and HI, and
therefore probes regions at lower densities
(e.g., Bergeron \& Stasi\'nska 1986).
At present, the evidence for the
connection between CIV absorbers and galaxies is still limited.
Here we examine the consequence of our model for CIV
absorption line systems, which may be tested by future observations.

 Figure 13 shows the column densisy ratio $\nciv/\nhi$
for photoionized clouds as a function of the ionization 
parameter $\Gamma$. Results are shown for the same cases
as shown in Fig.\ 3. In all cases, $\nciv$ increases rapidly
with $\Gamma$ and depends only weakly on the shape of $J(\nu)$
at $\Gamma\lsim 10^{-2}$. The flattening of $\nciv/\nhi$
at $\Gamma >10^{-2}$ for the hardest spectrum is due to the ionization
of CIV to CV. Comparing Fig.\ 13 with Fig.\ 3 we see 
immediately that, with respect to $\nhi$, $\nciv$ should be enhanced
in small halos (which have lower pressures).
Figure 14 shows $\nciv$ as a function of $D$
for halos with different $\vcir$ at $z=0.5$. 
The flux $J(\nu)$ is as specified in equation (18),
with $J_{-21}=0.1$, $\alpha=0.5$ and $\Theta (\nu\ge\nu_4)=0.1$.
Unlike HI and MgII
systems, the column density profile of CIV is quite flat in
large halos. The reason is that the density of CIV ions 
decreases rapidly to the inner
region of these halos, where
the pressure of gas is high and $\Gamma$ is low.
Thus, for a given halo $N_{\rm CIV}/N_{\rm MgII}$ should
increase with increasing impact parameter.
For the same reason, systems with high $\nciv$ are
associated with small halos with $\vcir\sim 150\kms$. 
To demonstrate this, we show in Figure 15 the impact parameter $D$,
at which $\nciv=10^{14}\cm^{-2}$ [corresponding to a
rest-frame equivalent width $W({\rm {CIV}}\lambda 1548)\sim 0.4\AA$
for unsaturated lines], as a function of $\vcir$ 
for one model as specified in the figure caption. 
For comparison, the curve for the MgII systems
in the same model (see the solid curve in Fig.10) is included.
In small halos, $D$ is much larger
for CIV systems than for MgII systems. In fact,
the cross-section weighted average of $\vcir$ is only 
about $100\kms$ for the CIV systems, but as large as
about $200\kms$ for MgII systems, in this particular model
with halo density given by the CDM spectrum. Our model thus
predicts that, while MgII systems are mostly associated with
bright galaxies, many CIV systems should be found 
to be associated with small galaxies where $\nmgii$ is low. 

The total number of CIV systems in our model is also shown in Fig.\ 12;
the ratio to the number of \mgii systems, and the evolution with
redshift, is similar to what is observed (e.g. Steidel \&
Sargent 1992). Unfortunately,
unlike those for LLSs and MgII systems, the number of CIV systems 
produced in small halos at high redshifts ($z\gsim 3$) is 
not negligible, because the number of such halos is large
at high redshift and CIV systems can arise in them owing to 
the low pressure. Thus the total number of CIV systems 
at high redshifts predicted by our model depends on whether or not 
the contribution from small halos are included.
In our calculation we have neglected the contribution from halos 
with $\vcir<50\kms$. The total number of CIV systems can 
increase by a factor of about two, if halos with $\vcir\ge 
30\kms$ are included. However, the existence of a cold phase in 
these small halos is uncertain, because the gas
there may not be able to cool owing to photoionization
heating (Efstathiou 1992; Steinmetz 1995). 
At $z\gsim 3$, the total number of CIV systems is smaller for
metallicity $Z=0.3\zsun$ than for $Z=0.1\zsun$, 
because the higher metallicity (which gives lower
pressure in a halo) leads to a large ionization parameter
($\Gamma\gsim 0.1$) in small halos so that 
the ratio between CIV column density and the total hydrogen column 
density {\it decreases} with increasing $\Gamma$ 
(see Fig.\ 13 where $\nciv/\nhi$ is shown).

Observations of QSO metal line systems at $z\sim 1$ have shown that a
large fraction ($\sim 70\%$) of metal line systems are ``CIV-only'' ones
(with rest-frame equivalent width $W\gsim 0.3\AA$) in which \mgii
systems (with the same equivalent-width limit) are not detected, while
there is only a small fraction of MgII systems where CIV is not
observed (Steidel \& Sargent 1992).
According to our model, the first result is understood mainly
from the fact that small halos have much larger
cross sections for CIV absorption than for MgII absorption (Fig.\ 15).
The second observational result is, however, not implied by our model
based on the photoionized phase,  
because systems in the most massive halos should show only MgII
absorption without detectable CIV, 
owing to the high pressures that are
prevalent everywhere within the cooling radius.
There are essentially two ways to produce CIV absorption in these
massive halos. The first is to have some gas in the cold phase in
regions of lower pressure, probably outside the cooling radius $r_c$.
Our simplified model assumes that no gas is present in the cold phase
at $r > r_c$, but in a realistic model the density of photoionized gas
would decline more smoothly with radius, which would allow for more CIV
to be produced in the outer regions.

  The other possible origin of CIV systems in massive halos is from
collisionally ionized gas. Although the gas in the hot phase is too hot
to produce detectable CIV (see \S 4.5), ``warm gas'' (with temperature
between the halo virial temperature and the
photoionization-equilibrium temperature),
which is in the process of cooling after being shocked 
or in heat conduction fronts between the hot and
the cold phase (e.g., Borkowski, Balbus, \& Fristrom 1990), should
give rise to CIV.
The fraction of carbon atoms that are in CIV in collisionally ionized
gas peaks at 0.3 at temperatures (1-2.5) $\times 10^5  \kelvin$.
Since the total hydrogen column density of the hot phase
at an impact parameter $D\lsim 30\kpc$ is typically
$10^{20}\cm ^{-2}$ for halos with $\vcir\sim 250\kms$ 
at $z\sim 1$, the typical CIV column density in the warm phase
is $\nciv\sim (f_wZ/\zsun)10^{16}\cm^{-2}$, where
$f_w$ is the total mass of warm gas in units of that of the
hot gas.
The maximum amount of gas that could be present in this warm phase
is obtained by assuming that a significant fraction of the cooling
energy from the hot phase is emitted in conduction fronts. This would
decrease the pressure of the hot phase, since the effective cooling
rate is increased.
Assuming pressure equilibrium, it then follows that 
$f_w\sim E_w\, \Lambda(T_h)T_w/\Lambda(T_w)T_h$, where $E_w$ is the
fraction of the cooling radiation emitted from the warm phase,
$\Lambda(T)$ is the cooling rate at temperature $T$, and $T_h$ and $T_w$
are the temperatures of the hot and the warm phases, respectively. 
To estimate $E_w$, we have calculated a simple plane-parallel,
stationary conduction front between temperatures $T_c=10^4\kelvin$
and $T_h = 2\times 10^6 \kelvin$ (see Zel'dovich \& Pickelner 1969).
This shows that the fraction of energy emitted at temperatures
(1-2.5) $\times 10^5 \kelvin$ can be at most 20\%. Thus, $f_w \ls
10^{-3}$, so the CIV column densities in massive halos from collisionally
ionized gas should only be $\sim 3\times  10^{12} \cm^{-2}$ (for $Z=0.3
\zsun$), suggesting that CIV is mostly produced in the photoionized
phase.

\subsection {The hot phase and highly ionized systems}

  For even more highly ionized species like OVI, 
our model also predicts
that they cannot be due to photoionized clouds in massive
galactic halos, because the ionization parameter is too low
to produce such ionization states.
The long dashed curve in Fig. 15 shows the impact
parameter at which $\novi=10^{14}\cm ^{-2}$.
It is clear that such OVI systems should not be commonly found
as photoionized clouds in halos with $\vcir\gsim 150\kms$.
The necessity for a collisionally ionized component to explain some of
the OVI lines has also been discussed in Giroux, Sutherland, \& Shull
(1994).
Some OVI might arise from an intermediate temperature phase as
discussed above, but the column densities are expected to be low, for
the same reason that they are low for CIV. 
Here we consider the possibility of producing highly ionized
species from the hot phase in our model.

  Given the density and temperature profiles (see equations 3 and 4)
of the hot phase, we can calculate the column densities of
different ions at any specific impact parameter.  
At the temperature of the hot phase (typically about $ 10^6$K
for galactic-sized halos), most of the gas is in highly ionized species,
and the strongest lines that are produced in the UV are caused by
lithium-like ions (e.g. CIV, NV, OVI, NeVIII and MgX).
Mulchaey et al. (1996) suggest that such highly ionized species 
in QSO absorption spectra can be used as a probe for a
hot medium in groups of galaxies. Here we examine 
whether these species have high enough column densities
in our model to be detected by observations.
As before we use the code CLOUDY to
calculate the relative column densities. The results 
of the column densities at a fixed impact parameter $D=20\kpch$ 
as a function of halo circular velocity are shown 
in Figure 16 for the species mentioned above.
(We note that the column densities depend only weakly on 
$D$ for $D=10$-30$\kpch$ in halos with $\vcir>150$.)
For comparison the HI column density given by the hot phase
is also plotted. The column densities of the metal species
are proportional to the metallicity of the hot gas, and   
we have assumed a metallicity of $0.1\zsun$ in our calculation. 
Since our results do not depend strongly on redshift, 
only those for halos at $z=2$ are shown.
   
  It is clear from Fig.\ 16 that strong absorption lines are 
not expected for any of these species. At the maximum column
densities shown in Fig.\ 16, the predicted rest-frame equivalent
width (in doublet) is about $0.1\AA$ for OVI and 0.01-0.02$\AA$ for 
the other metal species. Thus, while the equivalent width for
OVI is already close to the detection limit of HST, especially
when the metallicity is larger than $0.1\zsun$, the equivalent
widths for the other species are still too small to be
detectable. For halos at high redshifts, the predicted equivalent
width for OVI should be readily detectable by the
HIRAS echelle spectrograph on the Keck Telescope.
To test for the presence of the hot phase postulated in our model, high
resolution observations of OVI lines would be needed to
distinguish between gas in the photoionized and the hot phase, 
from the
width of the individual component lines. A wide, smooth component in
OVI absorption lines would be indicative of the hot phase, and would
also give invaluable information about the dynamical state of the hot
halo gas from a measurement of the hydrodynamic broadening. However,
these observations are difficult because the OVI lines are superposed
with the  $\lya$ forest.
Weymann (1995) suggests that the hot gas 
in galactic halos might also produce some weak broad 
Ly$\alpha$ lines. As we see from Fig.\ 16, the HI column 
density is extremely low in our model, because we have assumed 
that the hot gas in a halo has temperature similar to the
virial temperature of the halo.   

\section {Discussion and Summary}

  A model has been constructed where QSO metal absorption line systems
are produced in a halo of gas that was accreted into dark matter halos
and was shocked, forming a hot phase, and contains also a photoionized
phase in the form of clouds moving through the halo.
The clouds could either form from cooling of the diffuse hot gas, or
could form from inhomogeneities in the infalling material when the gas
is shocked, or be stripped from denser gas in satellite galaxies.
As they move through the hot gas falling in the gravitational potential
well, the clouds should accumulate in the centers of dark halos and
probably form gaseous disks and stars. The
absorption line systems we have modeled are caused by the clouds as they
move in towards the center, before they are aggregated on a forming
galaxy. Thus, the properties of the `halo gas' uncovered
by absorption line studies are essential for our 
understanding of galaxy formation. Equation (10) implies an 
average accretion rate of about several solar masses per year
for a typical galaxy with $\vcir \sim 200\kms$,
which in turn implies a total gas mass of about
$10^{10}$-$10^{11}\msun$ for these galaxies, consistent
with the mass of the disk of our galaxy.
The properies of galaxies formed in this way may depend
crucially on the state in which cold gas was accreted 
into the halos. In low-mass halos where the cooling time is short,
most of the accreted gas may merge directly to the center and dissipate 
its energy quickly, forming stars in cloud collisions,
and perhaps producing irregular galaxies. In more massive halos
where the cooling time is longer, accreted gas may be first 
incorportated into a hot halo and then cool to form 
cold clouds moving in a more regular way to the 
center, probably forming a disk. It is therefore important to 
see the differences in the properties of the absorption-line
systems around different galaxies. The cold gas accumulated in the
center of a halo may settle down to a rotationally supported gaseous
disk (or, in some cases, be partially supported as well by random motion
and thermal pressure), and produce the observed damped Lyman alpha
systems (see e.g., Wolfe 1995). The amount of gas observed
in such systems suggests that most of the gas that collapsed
in halos of $\vcir\gsim 50\kms$ at $z\sim 3$ should
stay in gaseous form (Mo \& Miralda-Escud\'e 1994;
Kauffmann \& Charlot 1994). Since damped Lyman alpha
systems at $z>2$ are observed to be metal poor 
($Z< 0.1 \zsun$, see Pettini et al. 1995),
the clouds in galactic halos at $z\gsim 2$
should also be metal poor. The `halo gas' must be more enriched
at lower redshifts to produce the observed metal line
systems with higher metallicity. It is unclear at present
how the gas in galactic halos was enriched. It is possible
that galactic halos at low redshifts contain much gas
that belonged previously to the interstellar medium of satellite
galaxies, or originated from supernova explosions during the onset
of star formation in the halo center.
 
  Much theoretical work remains to be done to see the plausibility
of our model and to make it more predictive, 
mainly to understand the physical processes determining the rate of
formation and destruction of gas clouds moving through the halo, the
distribution of cloud masses and velocities, etc. Many simplifications
have been necessary in this paper, but within the model uncertainties
we have found that the principal characteristics of LLSs and metal line
absorption systems can be reproduced by the accreted material
that forms the galaxies, in the form of clouds 
in pressure equilibrium with hot
halo gas, photoionized by the extragalactic background. The main results
are as follows:
 
(1) The masses of the clouds are in the range $10^5$-$10^6 \msun$,
when constrained to have a covering factor of about
unity along a line of sight in typical halos with circular 
velocity $\vcir \sim 200\kms$ and
within an impact parameter $\sim 30\kpch$,
as is observed for the QSO MgII systems.
Their typical sizes are near $1 \kpc$, and the neutral fractions are
$\sim 0.02$. Such clouds are stable against both gravitational
collapse and heat conduction.  

(2) The clouds in galactic halos can be the origin of most
absorption line systems with HI column densities $\nhi\gsim
10^{17} \cm^{-2}$ at redshifts $z\lsim 2$; however,
at $z\simeq 3$, the number of such systems 
predicted by our model is lower than observed. The number of absorption
systems can be brought into agreement with observations only if the
intensity of ionizing radiation is as low as allowed by the observations
of the number of quasars, and the fraction of mass accreting in halos
which is baryonic and in gaseous form is larger than $5\%$.

(3) For a typical halo with circular velocity
$\vcir\sim 200\kms$, the HI and MgII column densities 
decline rapidly with impact parameter as a result
of the pressure gradient in the halo.
CIV systems are more abundant at large radius because the higher
pressure in the inner parts of halos leads to lower ionization
states. 
At redshifts $z\sim 0.5$, the maximum impact parameter 
for producing a system with $\nhi\sim 10^{17}\cm ^{-2}$ is 
near $40\kpch$ in halos with $\vcir\gsim 200\kms$,
increasing slowly with $\vcir$; it decreases abruptly
at $\vcir\lsim 200\kms$. This result also applies to MgII systems 
with column density $\nmgii=10^{13}$-$10^{14}\cm ^{-2}$,
depending on the metallicity of these systems. The observed
cutoff of MgII systems at impact parameters $\gsim 40\kpch$
can arise as a result of the cooling radius and the pressure gradient
(through the effect on the neutral fraction in the clouds) of gas in
galactic halos.   

(4) Photoionized clouds should not be abundant
at large radius from massive halos, where the gas is virialized
but the cooling time is long. If photoionized clouds were present with
a large covering factor in regions where the cooling time is long,
massive clusters today would contain most of the Lyman limit systems,
which would show strong negative evolution due to the recent formation
of clusters. 

(5) Most LLSs and MgII systems at low redshifts are associated 
with massive halos in normal bright galaxies, while
most CIV systems should be found in the halos of
small galaxies and in the outer regions of big halos.

(6) The hot phase in galactic-sized halos can give rise to 
detectable absorption lines in OVI, while the column densities
predicted for other highly ionized species are low and difficult to
observe.

Further observations of absorption line systems and their relation to
galaxies should open a new era in our understanding of how galaxies
form, as the physical conditions of the gas in the halos around
galaxies, and the dependence of these conditions on galaxy luminosity
and morphology are unravelled.

\acknowledgments
  We thank John Bahcall, Cedric Lacey and Simon White for useful
discussions. We also thank Gary Ferland for providing us with the
photoionization code CLOUDY. JM is supported by the W. M. Keck
Foundation and by NASA grant NAG-51618, and HM by the
Ambrose Monell Foundation. We also
acknowledge support from NSF grant PHY94-07194 in the Institute for
Theoretical Physics in Santa Barbara.

\vfill
\eject

\begin{figure}
\epsscale{0.8}
\plotone{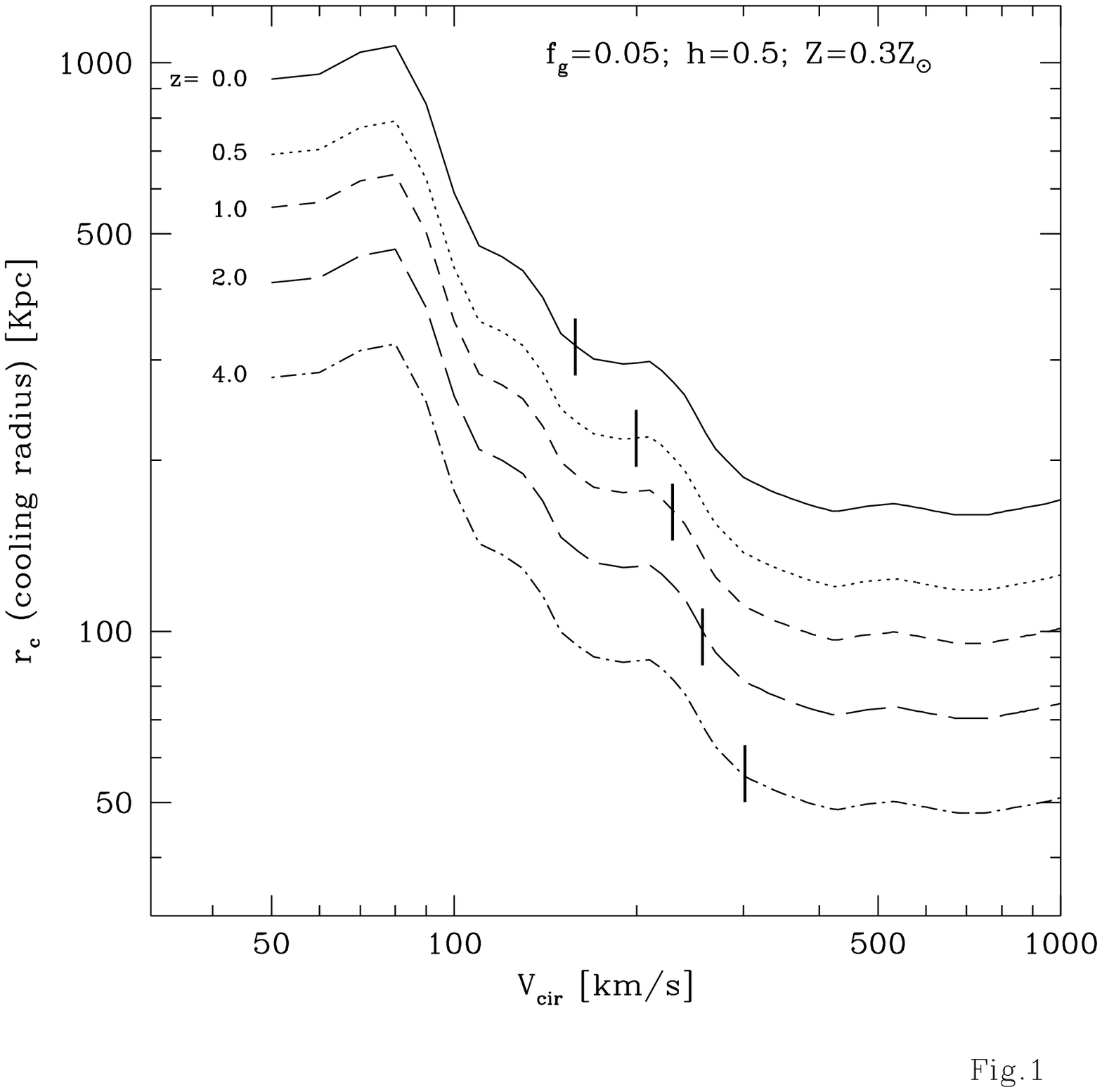}
\caption{
Cooling radius as a function of halo circular velocity,
shown at differemt redshifts $z$. The vertical marks 
show the value of $\vcir$ where the cooling radius is equal to the virial
radius. For lower $\vcir$, all the gas in the virialized
halo cools on a time shorter than the age of the halo, while at higher $\vcir$
the cooling time is longer. Result is show for a model with
$f_g=0.05$, $h=0.5$ and with metallicity $Z=0.3\zsun$. 
The cooling function
is adopted from Sutherland \& Dopita (1993).
}
\end{figure}
\begin{figure}
\epsscale{0.8}
\plotone{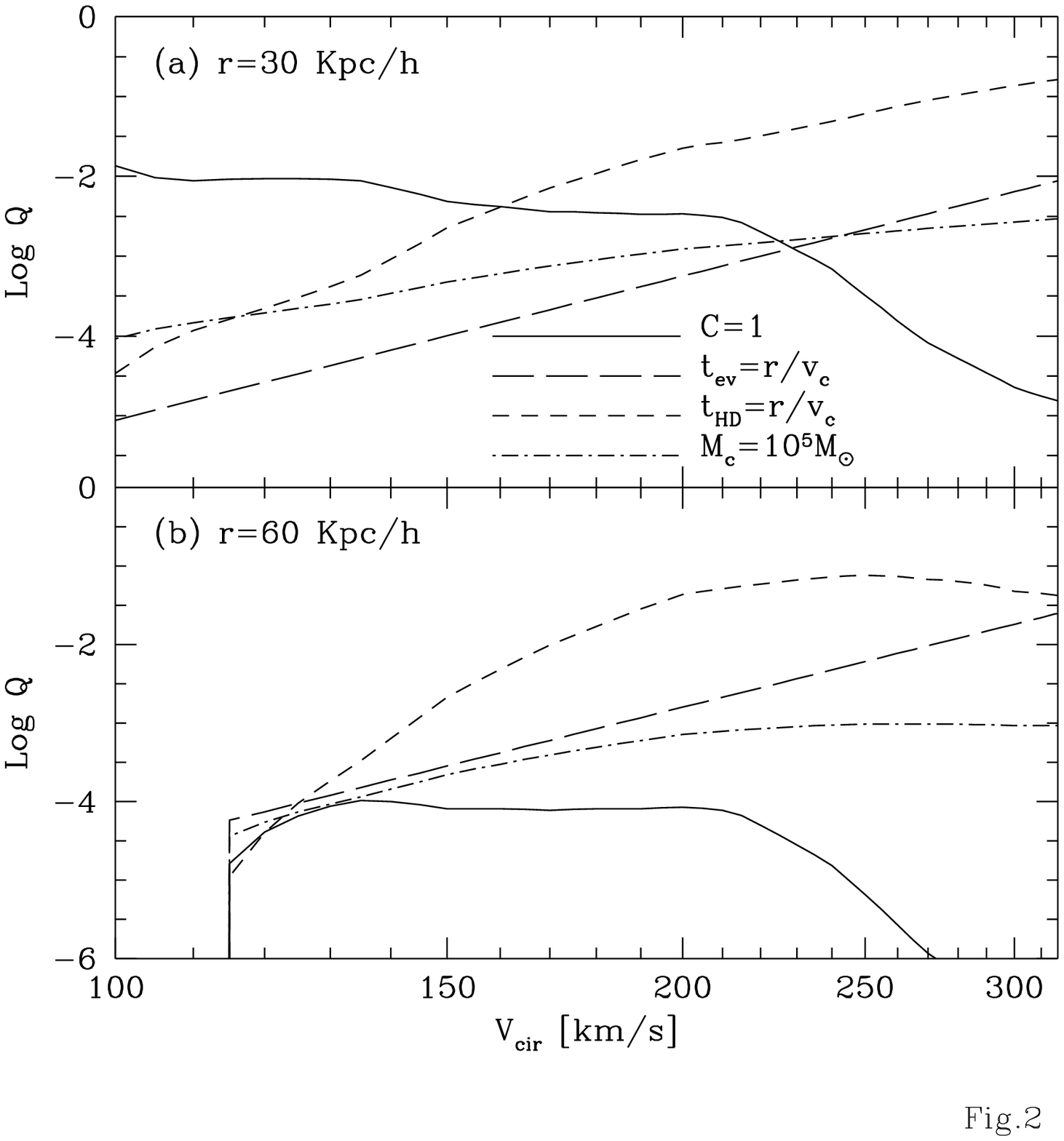}
\caption{
The allowed value of $Q$, defined as the ratio between the mass of cloud and
the critical mass for gravitational collapse,  
for clouds located at (a) $r=30\kpch$ and (b) $r=60\kpch$ from the centers
of halos with different $\vcir$. The solid curve shows the value of $Q$, 
below which the covering factor of clouds $C>1$. 
The long dashed (short dashed) curve
shows the limit of $Q$, above which the cloud will not 
be disrupted by heat conduction (hydrodynamic instability) 
as it moves through the halo over a radius $r$. 
The dot-dashed curve shows a constant cloud
mass: $M_c=10^5\msun$. 
Result is shown for halos at redshift $z=0.5$. We also 
assume $f_g=0.05$, $Z=0.3\zsun$ and $v_c=\vcir$.
}
\end{figure}
\begin{figure}
\epsscale{0.8}
\plotone{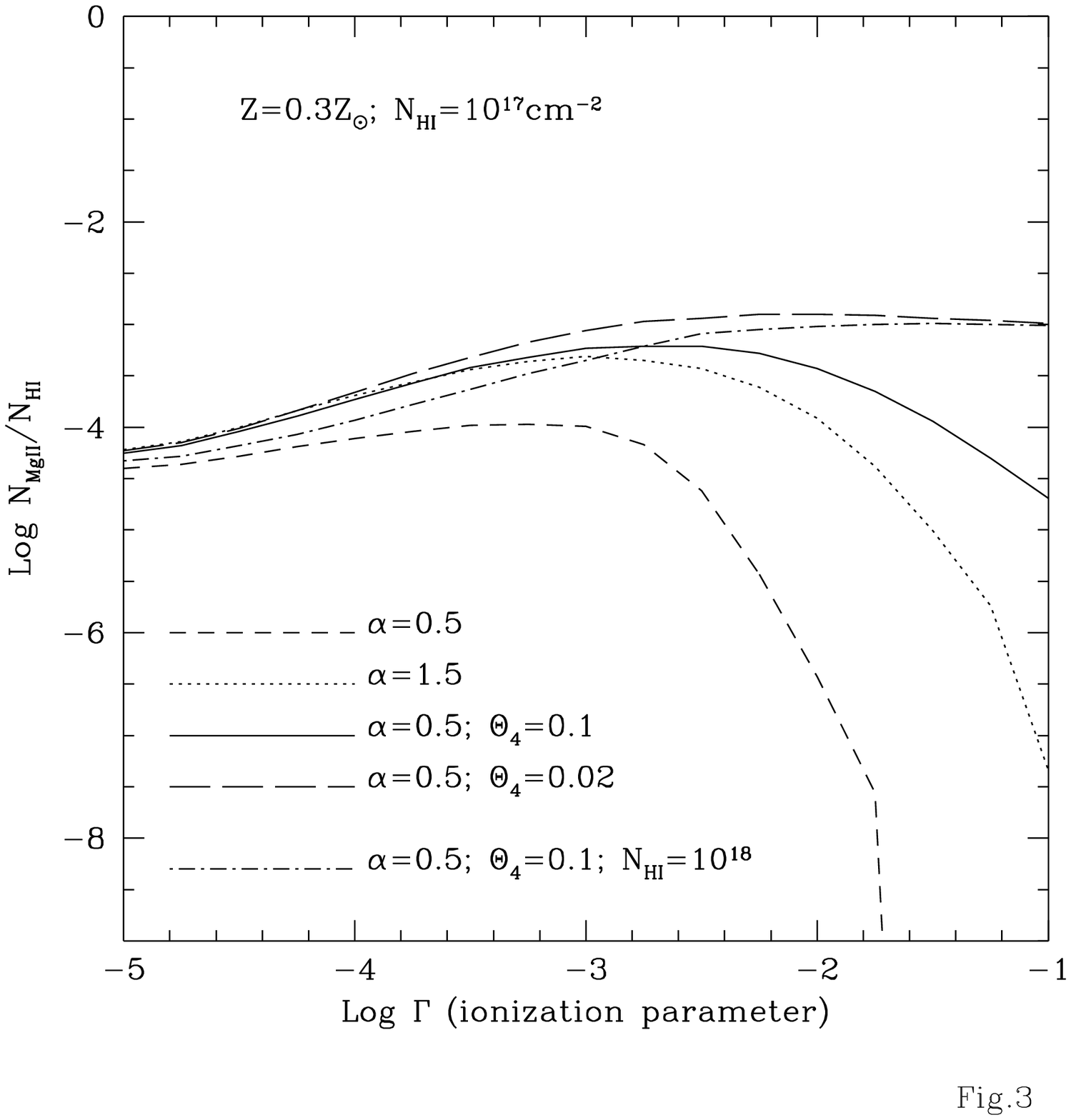}
\caption{
The ratio between MgII and HI column densities
as a function of the ionization parameter $\Gamma$ (defined as the 
ratio between the number density of photon above the
hydrogen Lyman limit frequency and the number density of total
hydrogen), for various shapes of the UV flux, as specified 
by the power index $\alpha$ and $\Theta_4\equiv \Theta(\nu\ge\nu_4)$. 
The dot-dashed curve shows the result for clouds
with $\nhi=10^{18}\cm ^{-2}$, while other curves
show results for clouds with $\nhi=10^{17}\cm^{-2}$. For all cases
a metallicity $Z=0.3\zsun$ is used.
}
\end{figure}
\begin{figure}
\epsscale{0.8}
\plotone{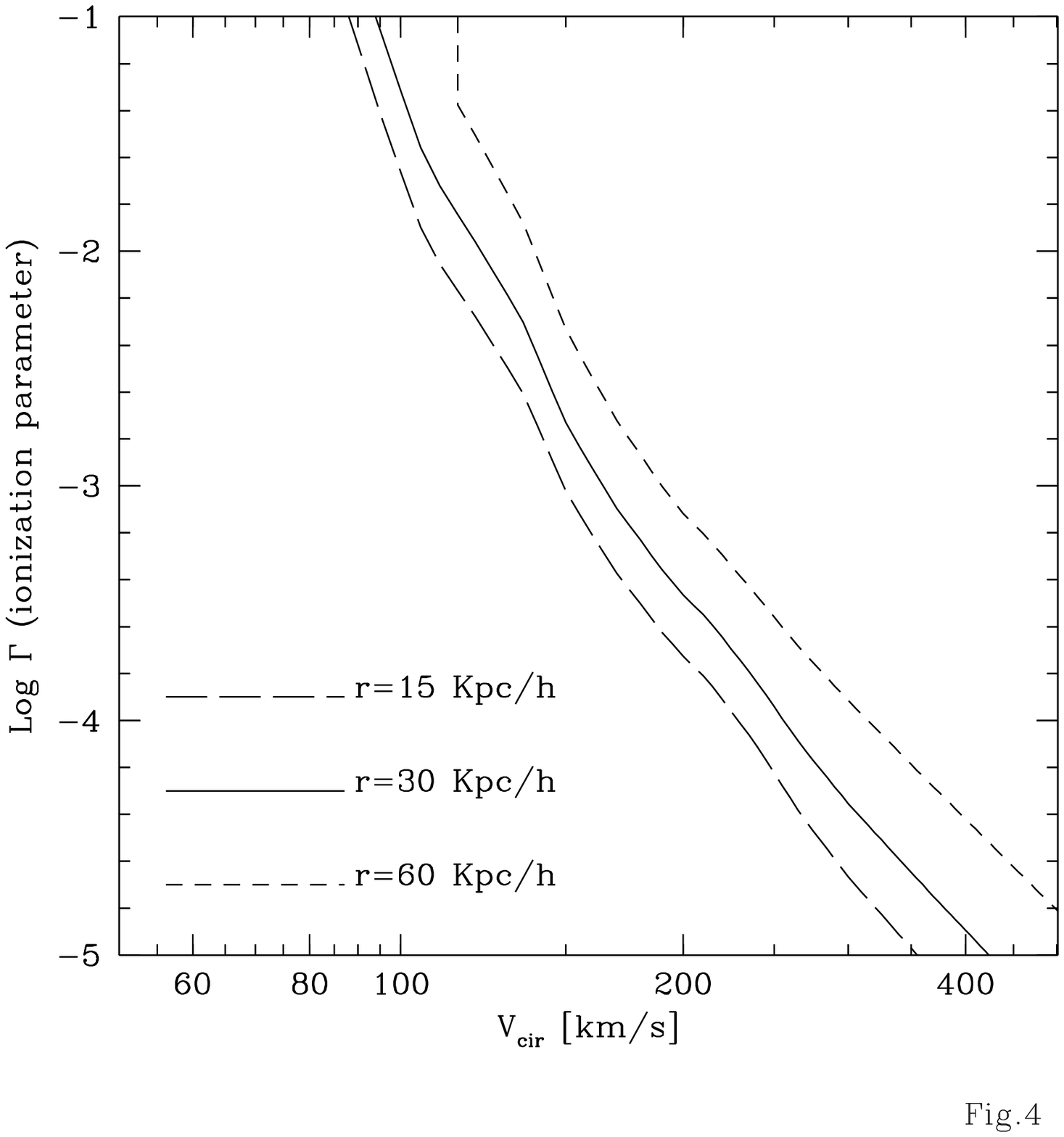}
\caption{
The ionization parameter $\Gamma$ of pressure confined clouds
located at a radius $r$ from halo center, as a function of halo
circular velocity. Results are shown for three typical radius
in halos at $z=0.5$, with $f_g=0.05$ and $Z=0.3\zsun$.
}
\end{figure}
\begin{figure}
\epsscale{0.8}
\plotone{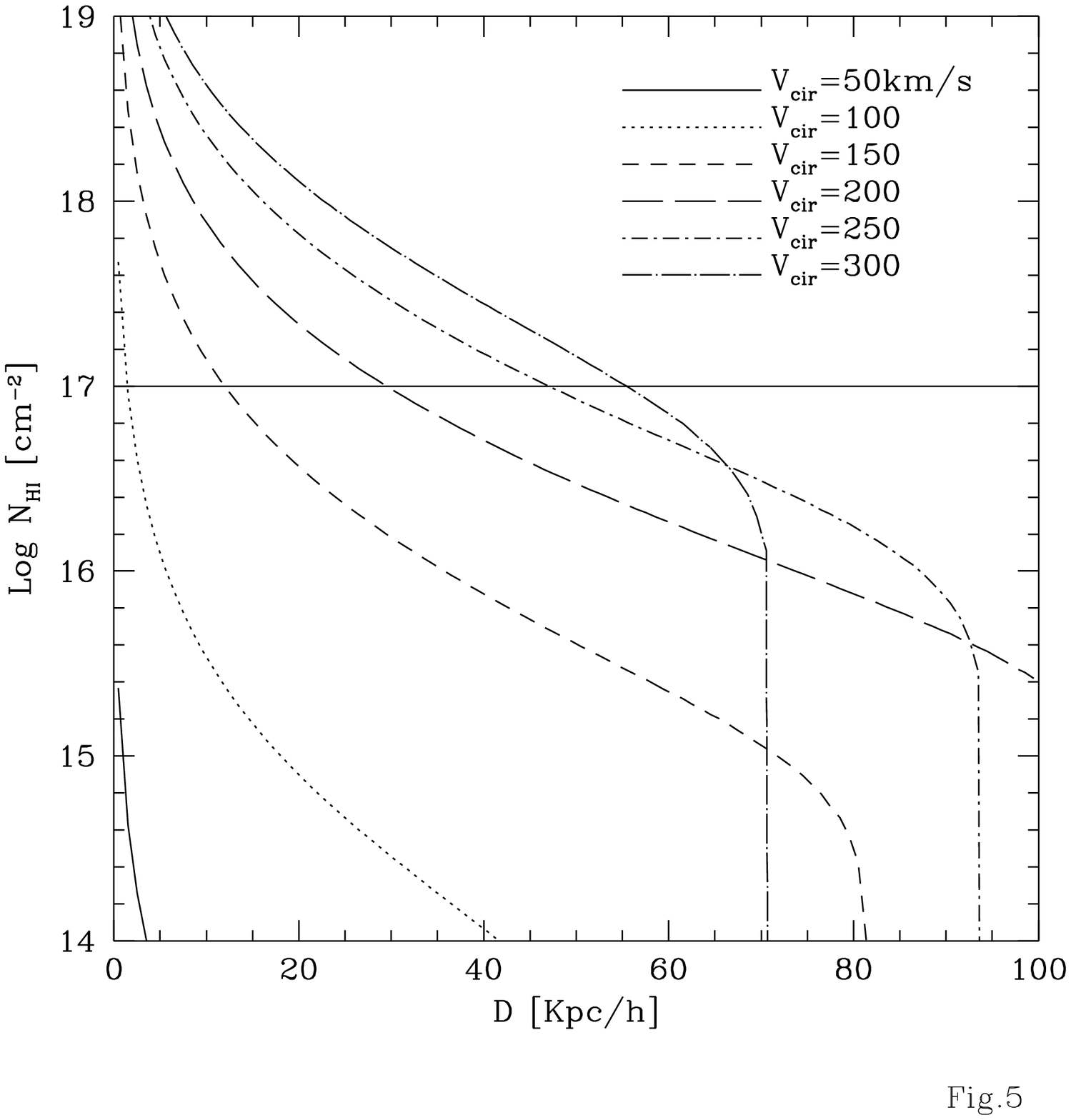}
\caption{
HI column density as a function of impact parameter $D$, for halos
at $z=0.5$ and with various $\vcir$. The model parameters 
are $f_g=0.05$, $Z=0.3\zsun$ and $v_c=\vcir$. The horizontal line
indicates $\nhi=10^{17}\cm ^{-2}$, above which a Lyman limit 
system is produced.                                                      
}
\end{figure}
\begin{figure}
\epsscale{0.8}
\plotone{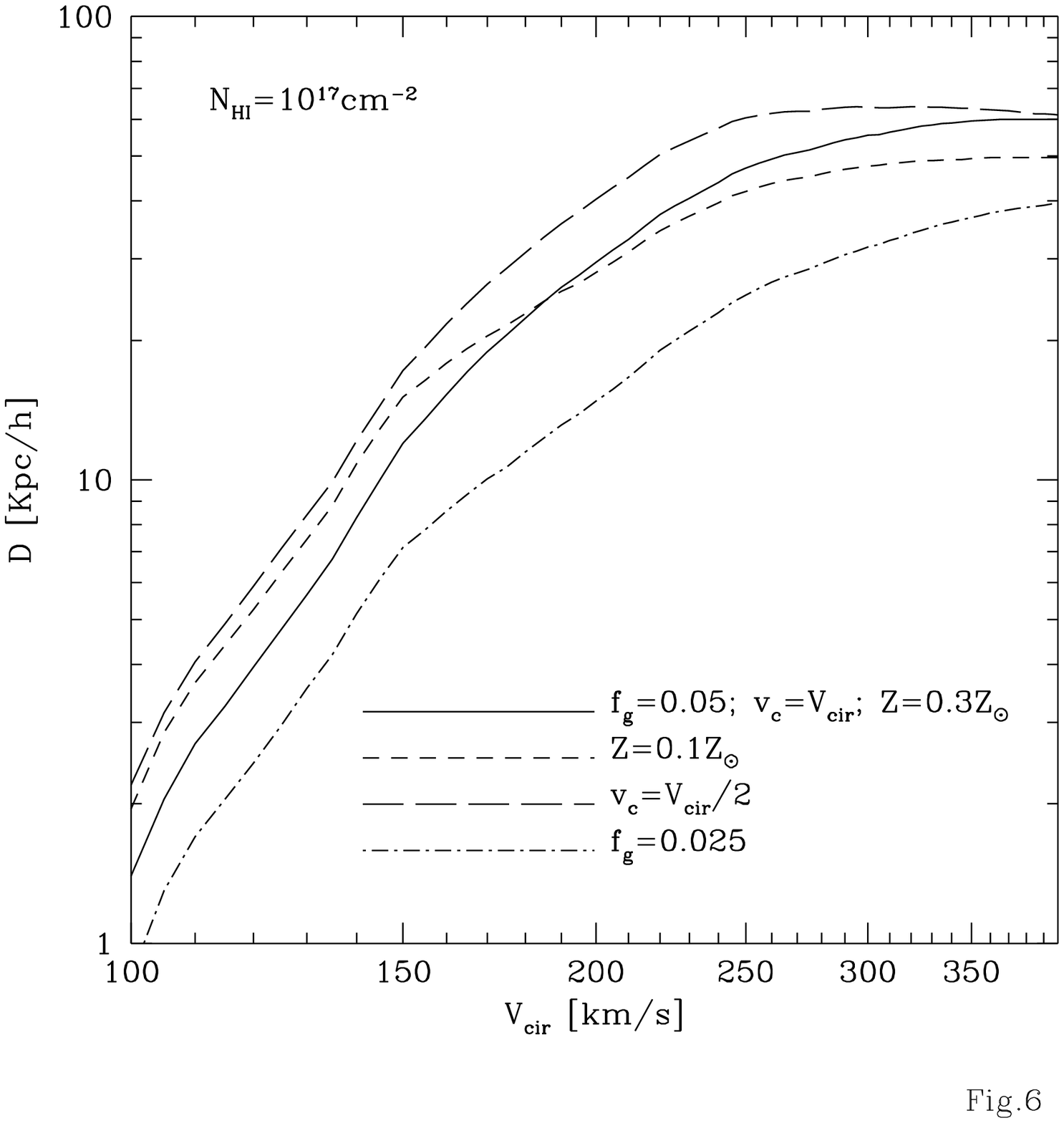}
\caption{
The impact paramter $D$, at which $\nhi=10^{17}\cm ^{-2}$, as a function
of halo circular velocity. 
The solid curve shows the result for
the `fiducial' model with $f_g=0.05$, $Z=0.3\zsun$ and $v_c=\vcir$; 
each other curve shows the result for a model in which one model parameter
differs from that for the fiducial model, as indicated in the panel.
Results are shown for halos at $z=0.5$. The UV flux has
$J_{-21}=0.1$, $\alpha=0.5$ and $\Theta_4=0.1$.
}
\end{figure}
\begin{figure}
\epsscale{0.8}
\plotone{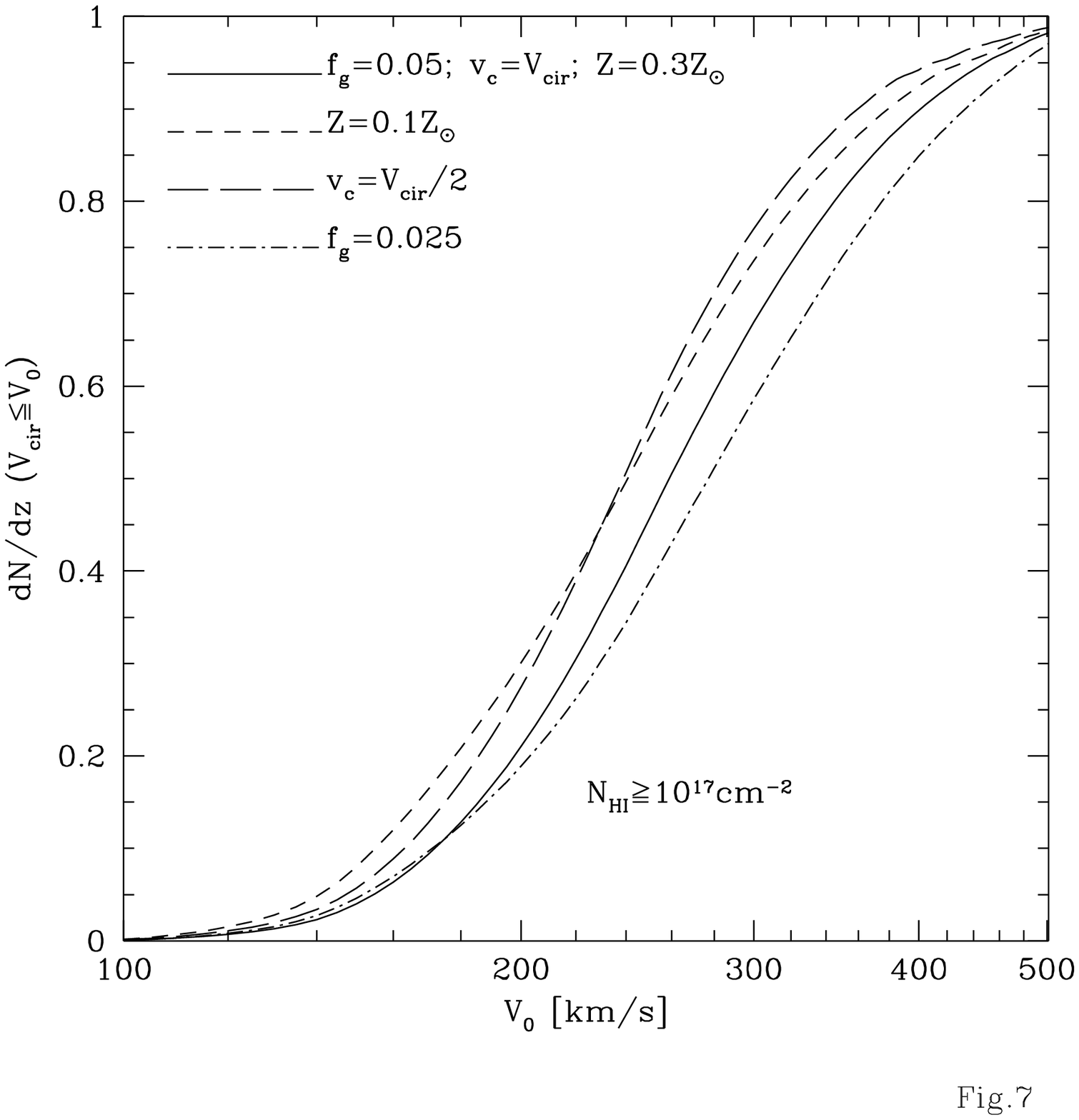}
\caption{
Cumulative cross section for absorption systems 
with $\nhi\ge 10^{17} \cm ^{-2}$ at $z=0.5$, given by halos with circular 
velocity $\vcir\le V_0$, as a function of $V_0$. The total 
cross section has been normalized to be 1. 
The results are shown for the same models as those in Figure 6.
}
\end{figure}
\begin{figure}
\epsscale{0.8}
\plotone{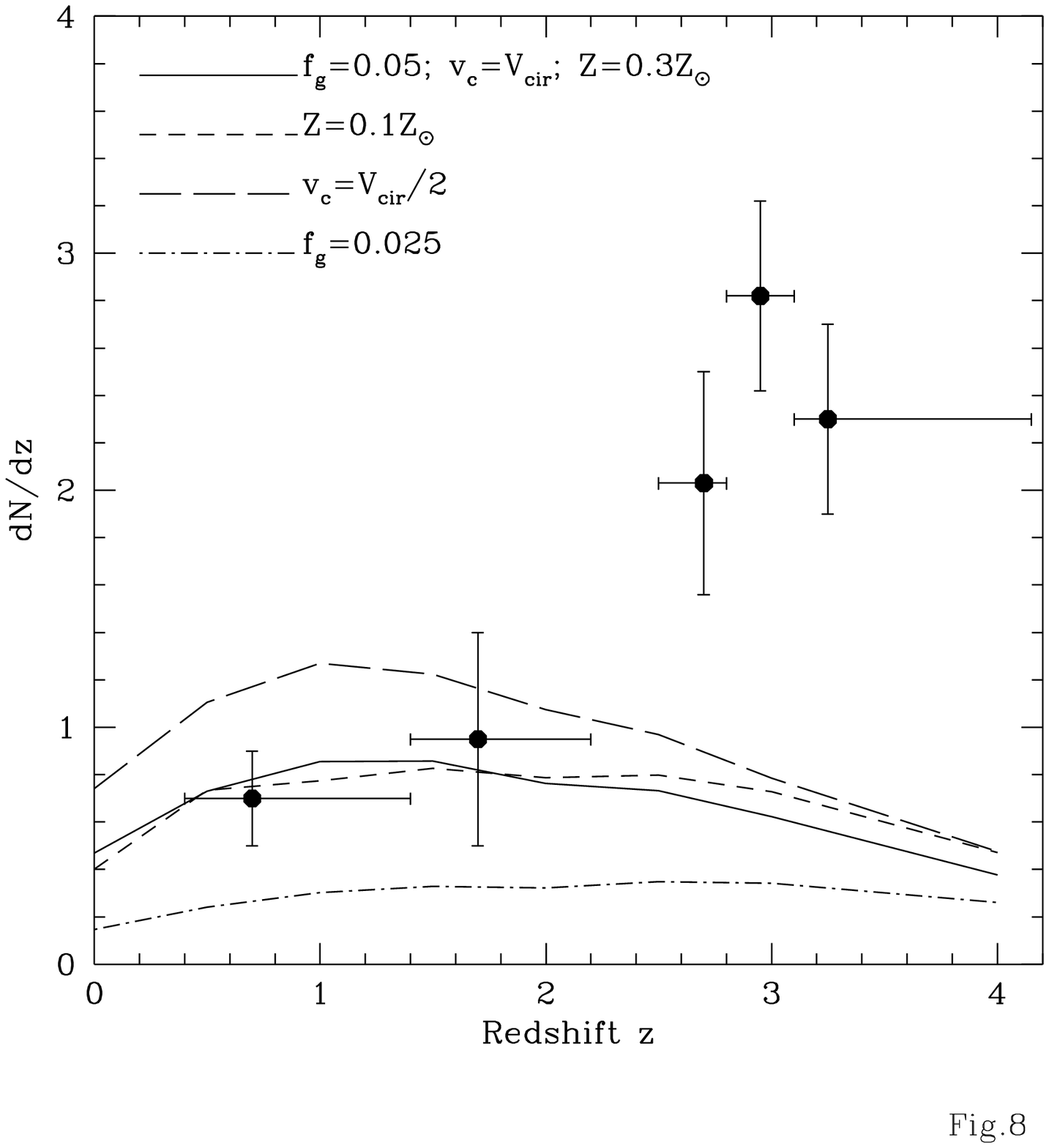}
\caption{
The number of $\nhi\ge 10^{17}\cm ^{-2}$ absorption systems per unit
redshift, as a function of redshift $z$. Curves show theoretical
models; data points show the observational result for
Lyman limit systems, adopted from Stengler-Larrea et al. (1995).
Results are shown for the same models as those in Figure 6.
The dependence of UV flux on $z$ is that specified in the text.
}
\end{figure}
\begin{figure}
\epsscale{0.8}
\plotone{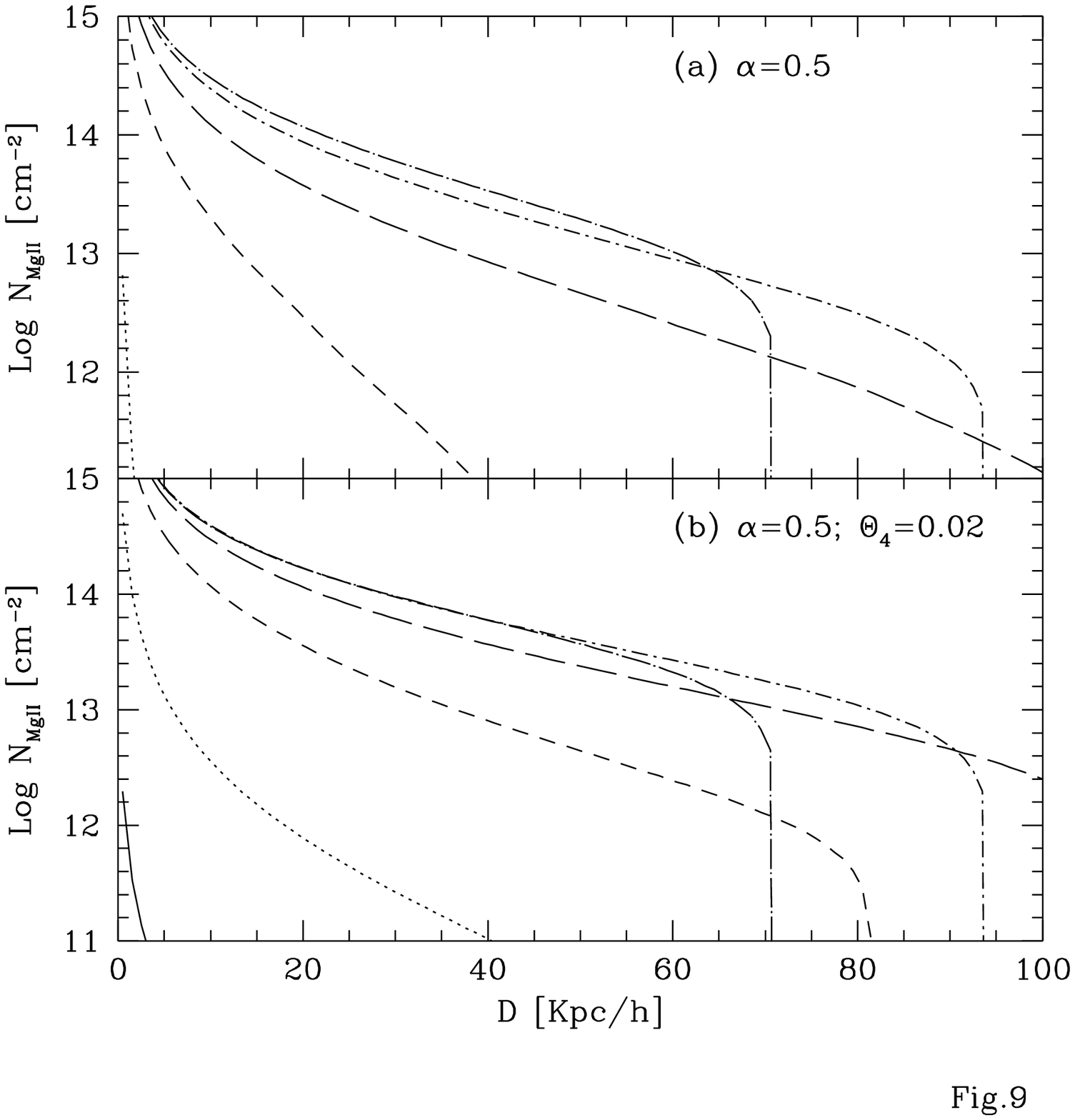}
\caption{
The MgII column density as a function of impact parameter, for halos
at $z=0.5$ and with circular velocity $\vcir=50$ (solid curve),
100 (dotted curve), 150 (short dashed curve), 200 (long dashed curve),
250 (dotted-short-dashed curve), $300\kms$ (dotted-long-dashed curve), 
in a model with $f_g=0.05$, $Z=0.3\zsun$ and $v_c=\vcir$.
Results are shown for two models of UV flux, one is a power law
with $\alpha=0.5$, the other has the same power index but with a
break at 4 Ryd so that $\Theta_4\equiv \Theta (\nu\ge \nu_4)=0.02$.
In both cases $J_{-21}=0.1$. 
}
\end{figure}
\begin{figure}
\epsscale{0.8}
\plotone{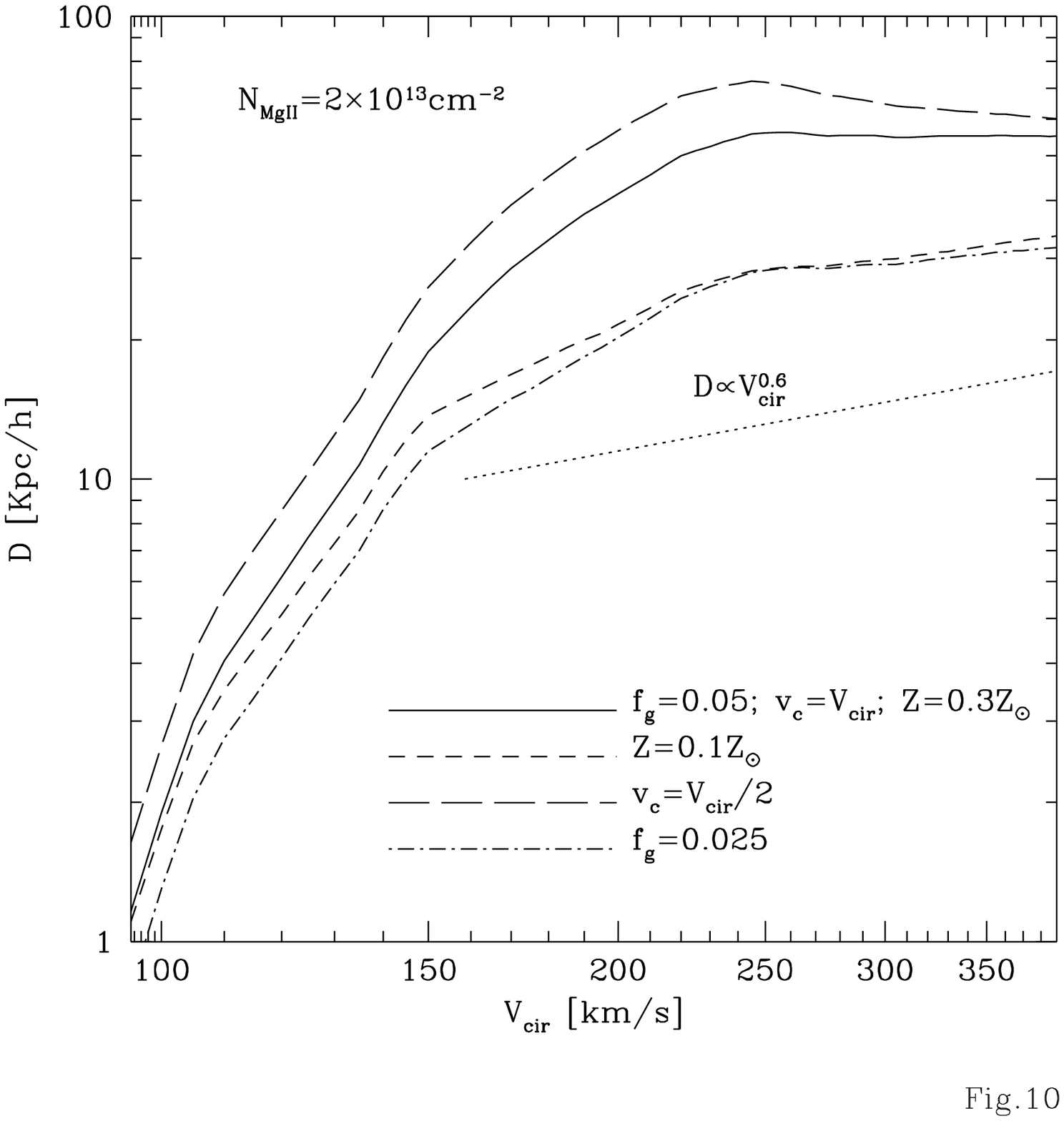}
\caption{
The impact parameter $D$, at which $\nmgii=2\times 10^{13}\cm ^{-2}$,
as a function of halo circular velocity. 
Results are show for halos at $z=0.5$, and for the same models
as shown in Figure 6.
The dotted line shows the relation 
$D\propto \vcir^{0.6}$, for the reason discussed in the text.
}
\end{figure}
\begin{figure}
\epsscale{0.8}
\plotone{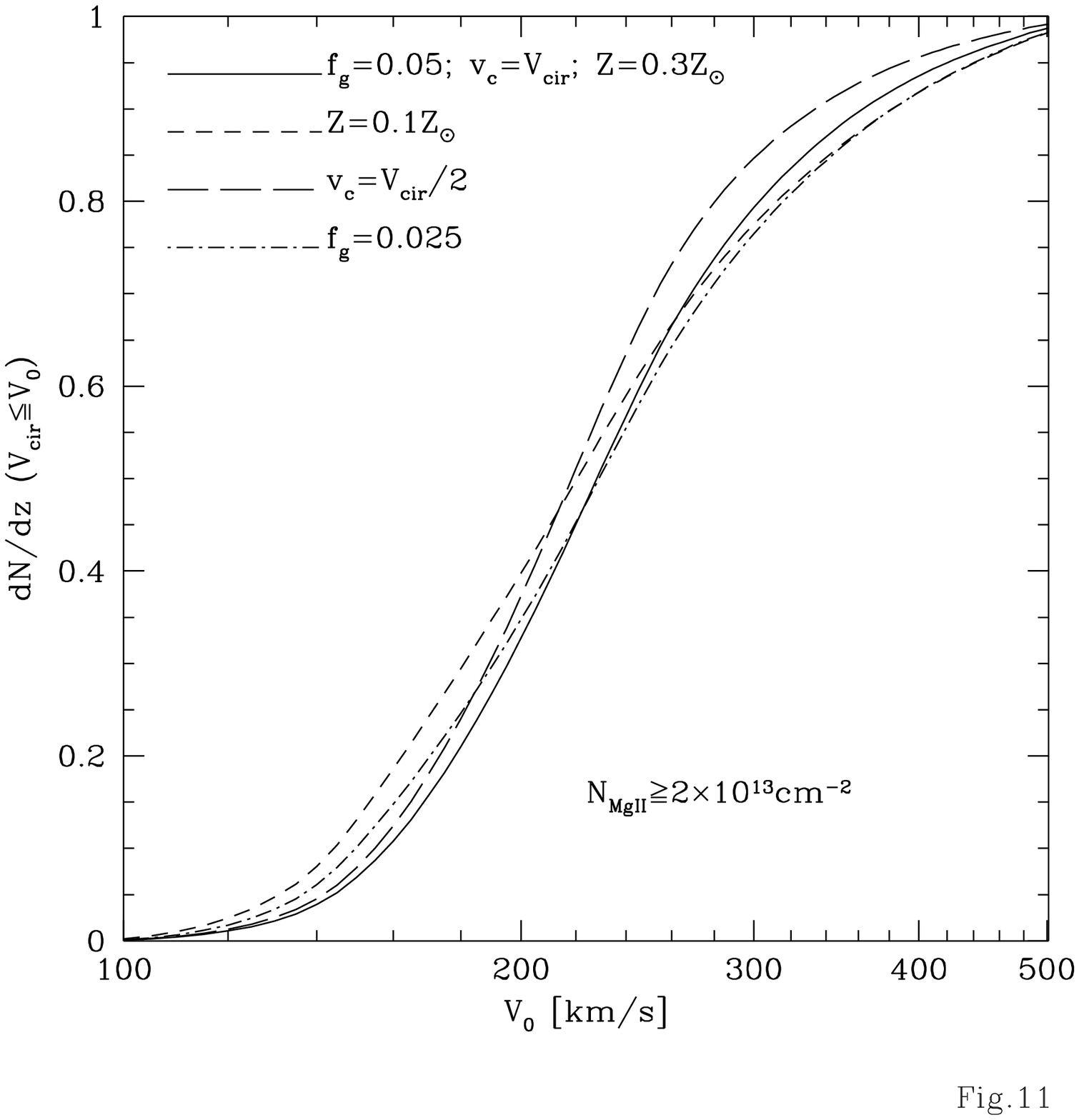}
\caption{
Cumulative cross section of MgII systems at $z=0.5$, with
$\nmgii\ge 2\times 10^{13}\cm ^{-2}$, given by halos with circular
velocity $\vcir\ge V_0$, as a function of $V_0$. Results
are shown for the same models as those in Figure 6.
}
\end{figure}
\begin{figure}
\epsscale{0.8}
\plotone{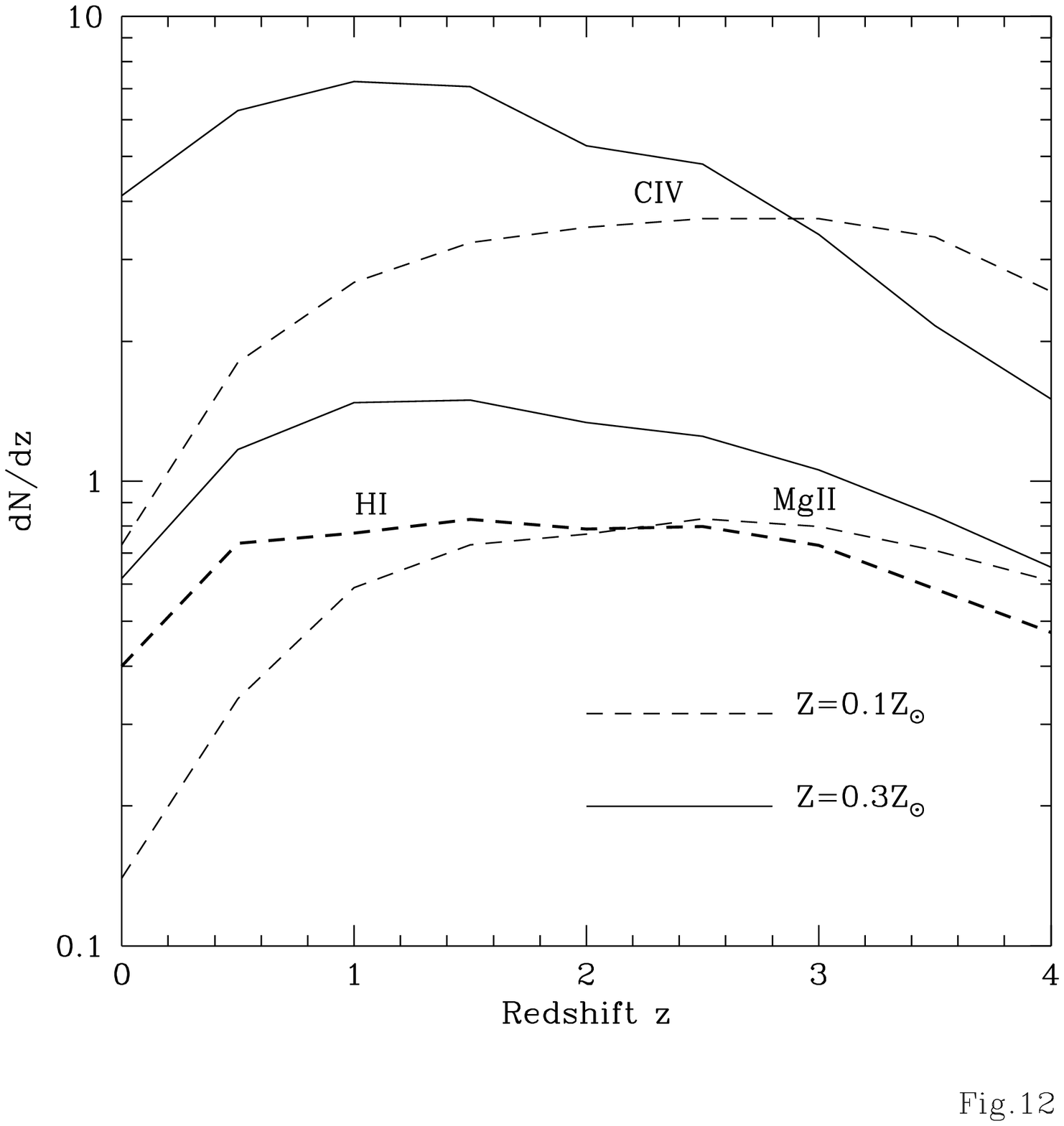}
\caption{
The number of CIV systems with $\nciv\ge 10^{14}\cm^{-2}$
and that of MgII systems with $\nmgii\ge 2\times 10^{13}\cm^{-2}$
per unit redshift, as a function of redshift $z$. 
Results are shown for two models. One (shown by the solid
curves) has $f_g=0.05$ and $v_c=\vcir$ with $Z=0.3\zsun$, wheras
the other (shown by the dashed curves) has the same $f_g$ and
$v_c$ but with $Z=0.1\zsun$. The dependence of the UV flux
on redshift is as specified in the text. For comparison
the result for the HI systems with $\nhi\ge 10^{17}\cm^{-2}$
in the latter model is shown by the thick dashed curve.
}
\end{figure}
\begin{figure}
\epsscale{0.8}
\plotone{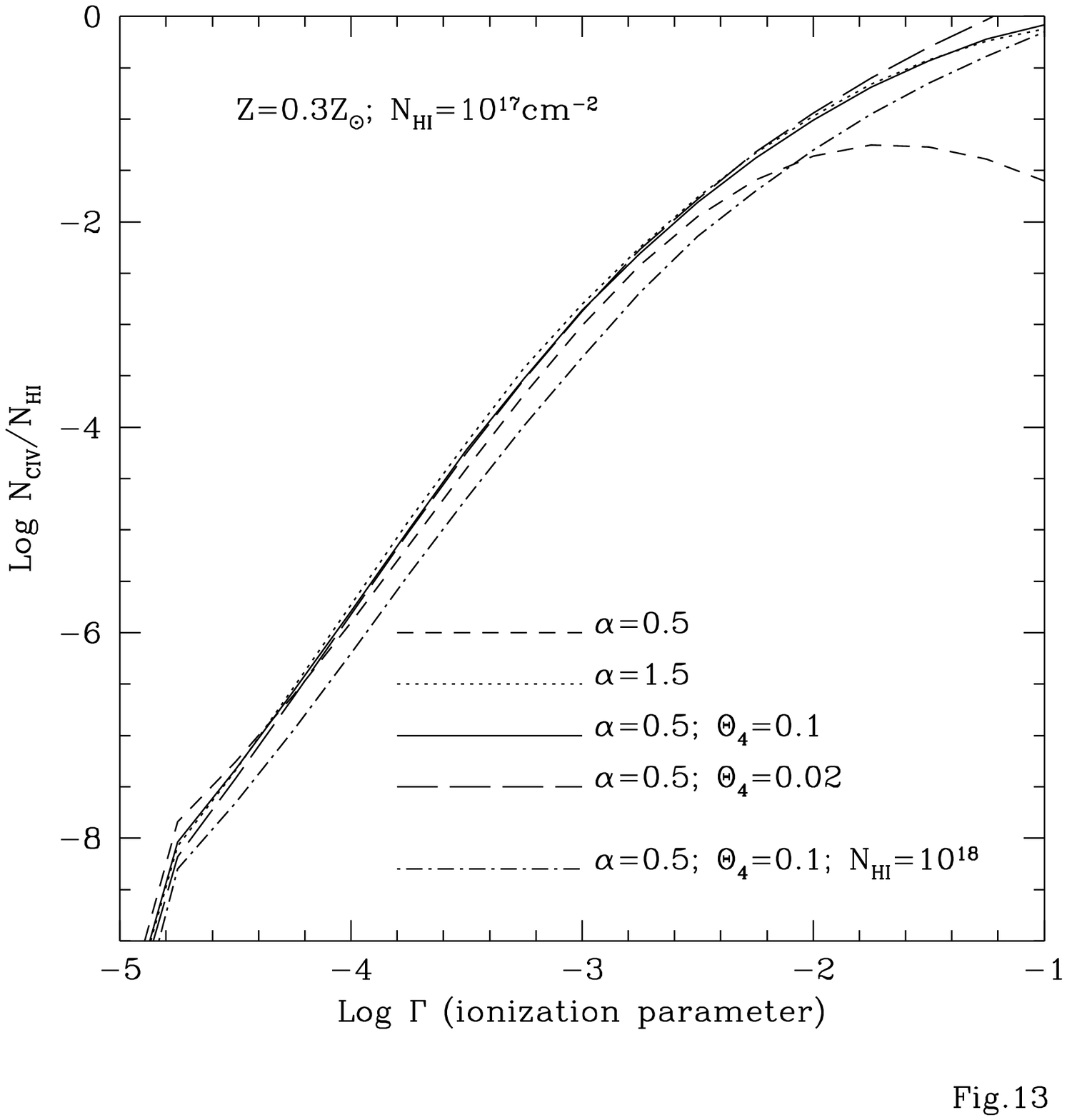}
\caption{
The ratio between CIV and HI comlun densities
as a function of the ionization parameter, for various  
shapes of UV flux. Results are shown for the same cases
as those shown in Figure 3.
}
\end{figure}
\begin{figure}
\epsscale{0.8}
\plotone{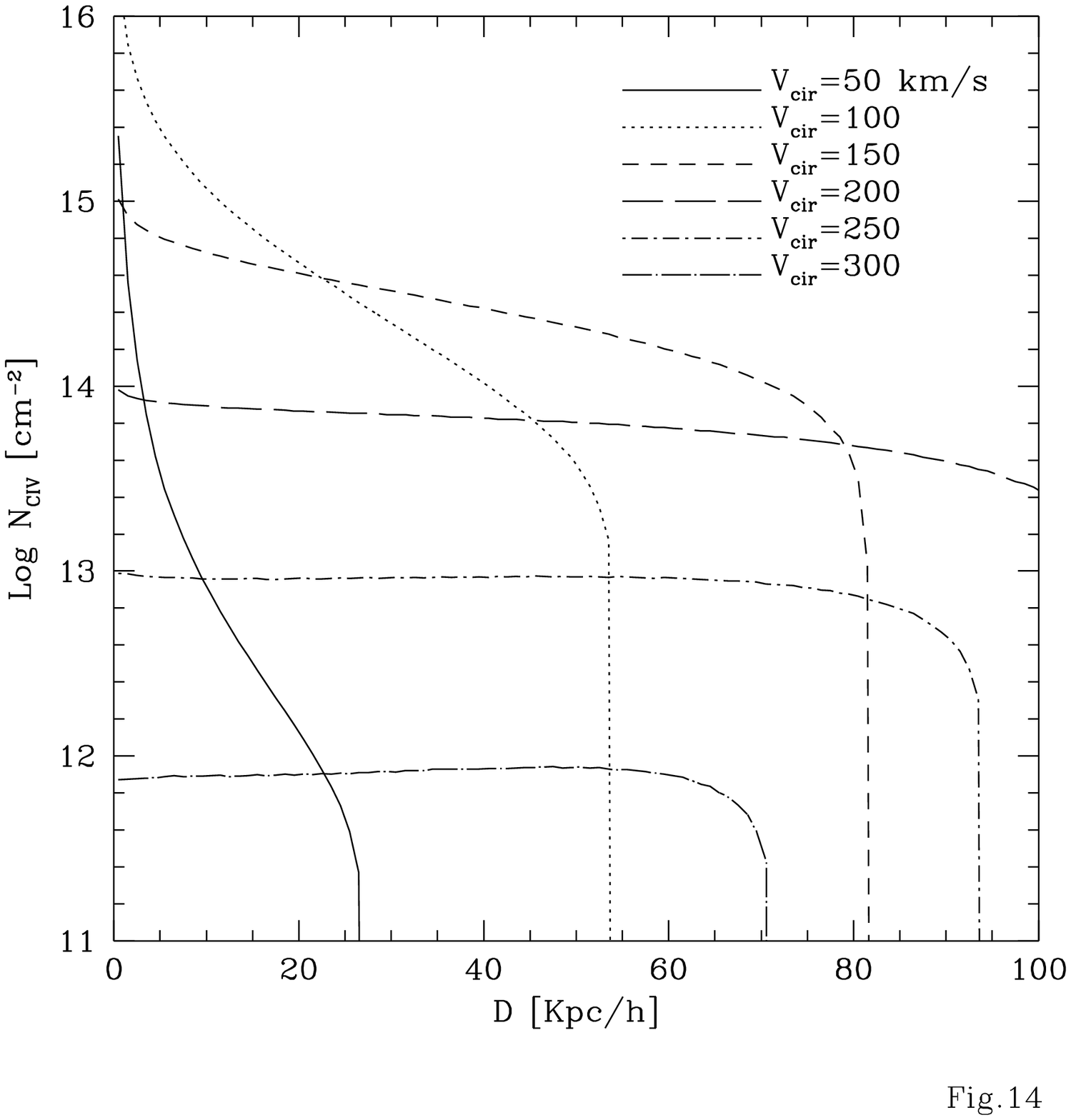}
\caption{
The CIV column density as a function of impact parameter,
for halos at $z=0.5$ and with different circular velocities,
in a model with $f_g=0.05$, $Z=0.3\zsun$ and $v_c=\vcir$.
The UV flux has $J_{-21}=0.1$, $\alpha=0.5$ and $\Theta_4=0.1$.
}
\end{figure}
\begin{figure}
\epsscale{0.8}
\plotone{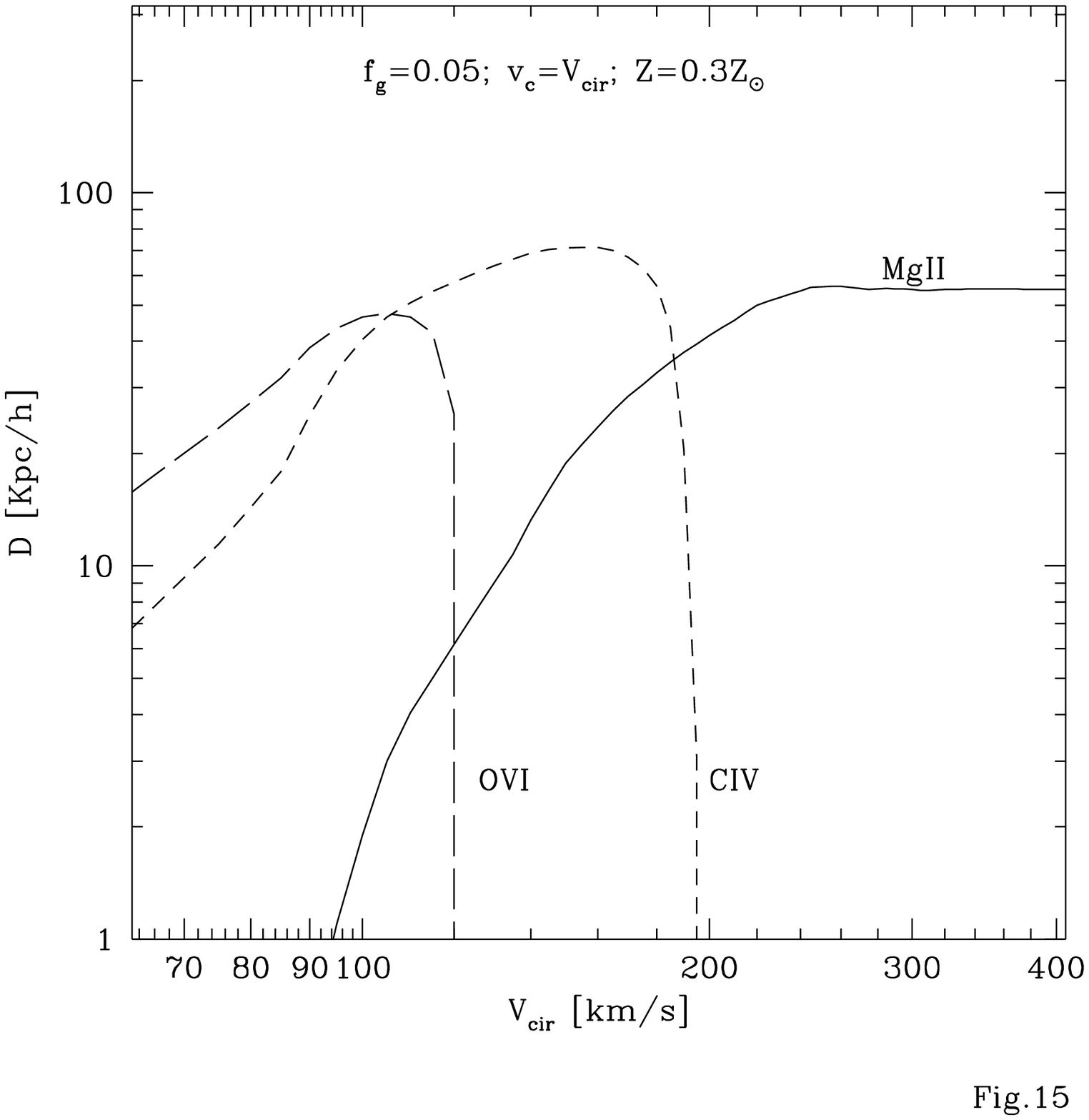}
\caption{
The impact parameters at which the CIV column density
$\nciv= 10^{14}\cm^{-2}$ (short dashed curve) and 
the OVI column density $\novi =10^{14} \cm ^{-2}$ (long dashed
curve), as a function of halo circular velocity $\vcir$.
Results are shown for halos at $z=0.5$,
in a model with $f_g=0.05$, $Z=0.3\zsun$ and $v_c=\vcir$.
The UV flux has $J_{-21}=0.1$, $\alpha=0.5$ and $\Theta_4=0.1$.
For comparison, the parameter at which 
$\nmgii=2\times 10^{13}\cm ^{-2}$
(shown in Figure 10) is also plotted here as the solid curve.
}
\end{figure}
\begin{figure}
\epsscale{0.8}
\plotone{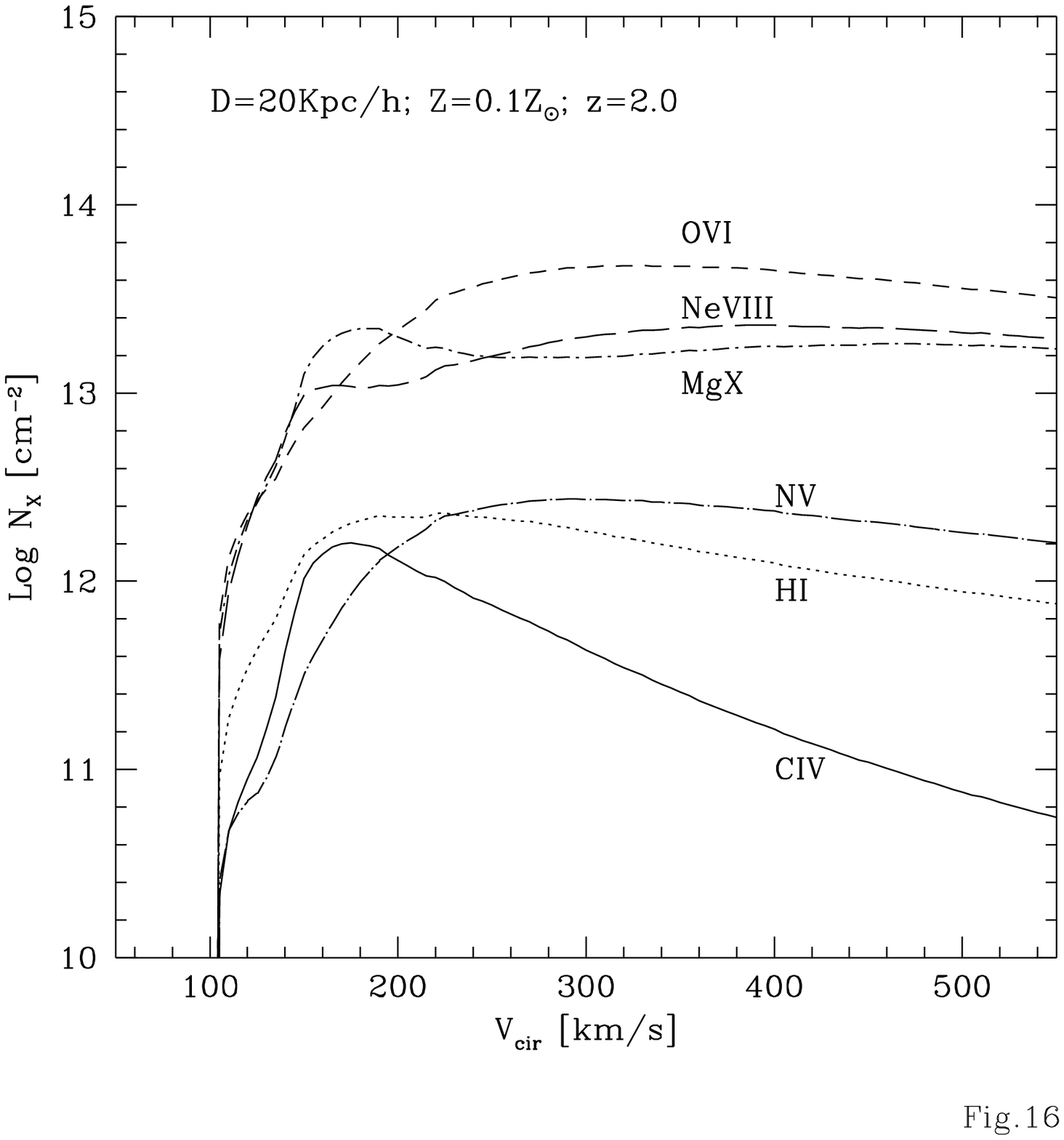}
\caption{
The column densities (at an fixed impact parameter $D=20\kpch$)
of different ions arising from the hot
phase in halos as a function of halo circular velocity $\vcir$.
Results are shown for halos at $z=2$. 
We have assumed $f_g=0.05$ and a metallicity
of $0.1\zsun$ for the hot gas.
}
\end{figure}

\begin{references}

\reference{o} Anninos P., Norman M., 1995, preprint 
\reference{o} Bahcall J.N., et al. 1995, ApJS, in press
\reference{o} Bahcall J.N., Spitzer L., 1969, ApJ, 156, L63
\reference{o} Balbus S.A., 1991, ApJ, 372, 25
\reference{o} Balbus S.A., Soker N., 1989, ApJ, 341, 611
\reference{o} Bardeen J.M., Bond J.R., Kaiser N., Szalay A.S., 1986, ApJ, 304, 15
\reference{o} Bechtold J., Ellingson E., 1992, ApJ, 396, 20
\reference{o} Begelmann M. C., McKee C.F., 1990, 358, 375
\reference{o} Bergeron J., 1995, in QSO Absorption Lines, ed. G. Maylan,
     Springer: Berlin, p.127
\reference{o} Bergeron J., Boiss\'e P., 1991, A\&A, 243, 344
\reference{o} Bergeron J., Cristiani S., Shaver P., 1992, A\&A, 257, 417
\reference{o} Bergeron J., Stasi\'nska G., 1986, A\&A, 169, 1
\reference{o} Borkowski K.J., Balbus S.A., Fristrom C.C., 1990, ApJ, 355, 501
\reference{o} Cen R.Y., Miralda-Escud\'e J., Ostriker J.P., Rauch M.R.,
     1994, ApJ, 437, L9 
\reference{o} Cowie L.L., McKee C.F., 1977, ApJ, 211, 135
\reference{o} Dean J.F., Davies R.D., 1975, MNRAS, 170, 503
\reference{o} Efstathiou G., 1992, MNRAS, 256, 43P
\reference{o} Evrard A.E., 1990, ApJ, 363, 349
\reference{o} Fabian A., 1994, ARA\&A, 32, 277
\reference{o} Fall M., Rees M.J., 1985, ApJ, 298, 18
\reference{o} Ferland G., 1993, University of Kentucky Department of Physics
     and Astronomy Internal Report 
\reference{o} Ferrara A., Field G.B., 1994, ApJ, 423, 665
\reference{o} Giroux, M. L., Sutherland, R. S., \& Shull, M. J. 1994, ApJ, 435, L97
\reference{o} Gunn J., 1982, in Astrophysical Cosmology, ed. H.A. Br\"uck,
     G.V. Coyne \& M.S. Longair, Pontifical Academy of Sciences:
     Vatican, p233   
\reference{o} Haardt F., Madau P., 1995, submitted to ApJ
\reference{o} Haehnelt M., Steinmetz M., Rauch M., 1996, ApJ, submitted
\reference{o} Hernquist L., Katz N., Weinberg D.H., Miralda-Escud\'e J., 
     1996, ApJ, 457, L51 
\reference{o} Jakobsen P., Boksenberg A., Deharveng J.M., Greenfield P.,
     Jedrzejewski R., Paresce F., 1994, Nat, 370, 35
\reference{o} Kang, H., Shapiro, P. R., Fall, S. M., \& Rees, M. J. 1990, 
     363, {488}
\reference{o} Katz N., Weinberg D.H., Hernquist L., Miralda-Escud\'e J., 1996,
     ApJ, 457, L57
\reference{o} Kauffmann G., Charlot S., 1994, ApJ, 430, L97 
\reference{o} Lacey C., Cole S., 1993, MNRAS, 262, 627
\reference{o} Loewenstein M., 1989, MNRAS, 238, 15
\reference{o} Loewenstein M., 1990, ApJ, 349, 471
\reference{o} Madau P., 1992, ApJ, 389, L1
\reference{o} Managoli A., Rosner R., Fryxell B., 1990, MNRAS, 247, 367
\reference{o} McCammon D., Sanders W.T., 1984, ApJ, 287, 167
\reference{o} McKee C.F., Cowie L.L., 1977, ApJ, 215, 213
\reference{o} Miralda-Escud\'e J., Cen R.Y., Ostriker J.P., Rauch M., 
     1995, submitted to ApJ
\reference{o} Miralda-Escud\'e J., Ostriker J.P., 1990, ApJ, 350, 1
\reference{o} Mo H.J., 1994, MNRAS, 269, L49
\reference{o} Mo H.J., Miralda-Escud\'e J., 1994, ApJ, 430, L25
\reference{o} Mo H.J., Miralda-Escud\'e J., Rees M.J., 1993, MNRAS, 264, 705
\reference{o} Mulchaey J., Mushotzky R.F., Burstein D., Davis D.S., 1996,
     ApJ, 456, L1
\reference{o} Murray S.D., White S.D.M., Blondin J.M., Lin D.N.C., 1993, ApJ, 407,
     588
\reference{o} Navaro J., Frenk C., White S.D.M., 1995, MNRAS, in press
\reference{o} Nulsen P.E.J., 1986, MNRAS, 221, 377
\reference{o} Peebles P.J.E., 1980, The Large-Scale Structure of the Universe,
     Princeton University Press: Princeton
\reference{o} Pettini M., King D.L., Smith L.J., Hunstead R.W., 1995,
     in QSO Absorption Lines, ed. G. Meylan, Springer: Berlin, p.71
\reference{o} Petitjean P., Bergeron J., 1990, A\&A, 231, 309
\reference{o} Petitjean P., Bergeron J., Puget J.L., 1992, A\&A, 265, 375
\reference{o} Press W.H., Schechter P., 1974, ApJ, 187, 425
\reference{o} Sargent W.L.W., Steidel C.C., Boksenberg A., 1989, ApJ, 334, 22
\reference{o} Spitzer L., 1978, Physical Processes in the Interstellar Medium,
     Wiley: New York
\reference{o} Steidel C.C., 1990, ApJS, 74, 37
\reference{o} Steidel C.C., 1993, in The Environment and Evolution of Galaxies,
     eds. J.M. Shull and H.A. Thronson, Kluwer: Dordrecht, p.263
\reference{o} Steidel C.C., 1995, 
     in QSO Absorption Lines, ed. G. Meylan, Springer: Berlin, p.137
\reference{o} Steidel C.C., Sargent W.L.W., 1992, ApJS, 80, 1
\reference{o} Steinmetz M., 1995, preprint
\reference{o} Stengler-Larrea E.A., et al. 1995, ApJ, 444, 64
\reference{o} Storrie-Lombardi L.J., McMahon R.G., Irwin M.J., Hazard C.,
     1994, ApJ, 427, L13
\reference{o} Sutherland R.S., Dopita M.A., 1993, ApJS, 88, 253
\reference{o} Tabor G., Binney J., 1993, MNRAS, 263, 323
\reference{o} Tenorio-Tagle G., Rozyczka M., Bodenheimer P., 1990,
     A\&A, 237, 207
\reference{o} Vogel S., Reimers D., 1995, A\&A, 294, 377
\reference{o} Walker T.P., Steigman G., Schramm D.N., Olive K.A., Kang H-S.,
     1991, ApJ, 376, 51 
\reference{o} Wang Q., McCray R., 1993, ApJ, 409, L37
\reference{o} Waxman E.,  Miralda-Escud\'e J., 1995, ApJ, 451, 451
\reference{o} Weymann R.J., 1995, 
     in QSO Absorption Lines, ed. G. Meylan, Springer: Berlin, p.3
\reference{o} White S.D.M., 1995, in 1993 Les Houches Lectures,
     ed. R. Schaeffer, in press
\reference{o} White S.D.M., Rees M.J., 1978, MNRAS, 183, 341
\reference{o} Wolfe A., 1995,
     in QSO Absorption Lines, ed. G. Meylan, Springer: Berlin, p.13
\reference{o} Yanny B., York D.G., Williams T.B., 1990, ApJ, 351, 377
\reference{o} Zel'dovich Ya. B., Pikel'ner S.B., 1969, Soviet Phys.
JETP, 29, 170 
\end{references}
\end{document}